\documentclass[twocolumn,showpacs,prd,floatfix,axodraw]{revtex4}
\usepackage{mathrsfs}
\usepackage{graphicx,,booktabs,bm}
\usepackage{overpic}
\usepackage{color}
\usepackage{amssymb}

\usepackage{mathrsfs,bm,amsmath,amssymb}
\usepackage{longtable,lscape}
\usepackage{txfonts}
\usepackage{amssymb}
\usepackage{indentfirst}
\usepackage{graphicx,,booktabs}
\usepackage{multirow,ulem}

\usepackage{color}
\usepackage{amssymb}

\definecolor{cover}{rgb}{0.77,0.87,0.88}
\definecolor{blueone}{rgb}{0.1,0.1,.7}
\definecolor{citec}{rgb}{0.14,0.47,0.09}
\definecolor{two}{rgb}{0.0,0.5,0.}
\definecolor{three}{rgb}{.5,.1,0.15}
\usepackage[bookmarks=true,bookmarksopen=false,plainpages=false,breaklinks=true,
   bookmarksnumbered=true,hypertexnames=false,
   filecolor=blue,urlcolor=three,menucolor=three,
   linkcolor=three,citecolor=blueone, colorlinks,
   anchorcolor=blue,runcolor=pink,frenchlinks=red
   pdfstartview=FitH,pdftitle=title,%
   pdfauthor=author]{hyperref}

\begin{document}
\title{The possible $K^{*}\Sigma^{*}$ molecular state}
\author{Yin Huang$^{1}$\footnote{corresponding author}} \email{huangy2019@swjtu.edu.cn}
\author{Dan Jiang$^{1}$}
\author{Feng Zhang$^{2}$}
\author{Bo Nan Zhang$^{3}$\footnote{corresponding author}} \email{bnzhang@imnu.edu.cn}
\affiliation{$^{1}$School of Physical Science and Technology, Southwest Jiaotong University, Chengdu 610031,China}
\affiliation{$^{2}$School of Physics and Electric Engineering, Weinan Normal University, Weinan 714099, China}
\affiliation{$^{3}$College of Physics and Electronic Information, Inner Mongolia Normal University,Hohhot, 010022, China}

\begin{abstract}
Within the framework of the one-boson-exchange model, we systematically investigate the interaction between the vector meson $K^{*}$ and the baryon $\Sigma^{*}$ with the aim of exploring the possibility of forming hadronic molecular states. The $K^{*}\Sigma^{*}$ interaction potential is constructed from $\rho$, $\omega$, and $\pi$ meson exchanges, and the nonrelativistic Schr\"odinger equation is solved using the Gaussian expansion method. The binding energies are calculated for different total angular momenta $J^{P}$ and isospin channels $I=1/2$ and $I=3/2$.  Our results show that $S$--$D$ wave mixed $K^{*}\Sigma^{*}$ molecular states with $J^{P}=1/2^{-}$ can be formed only in the $I=3/2$ channel, while no bound state appears in the $I=1/2$ channel. In addition, the $S$--$D$ wave mixed states with $J^{P}=3/2^{-}$ and $J^{P}=5/2^{-}$ are also found to support bound-state solutions.
For higher partial-wave states in our study, the binding mechanism mainly arises from the interplay between partial-wave mixing and non-central interactions. In particular, the $J^{P}=1/2^{+}$ channel does not support a bound state, as the meson-exchange interaction is predominantly repulsive. Our analysis further supports the interpretation of the experimentally observed $N(2250)$ and $\Delta(2200)$ states as $K^{*}\Sigma^{*}$ molecular candidates, corresponding to $I=1/2,\ J^{P}=9/2^{-}$ and $I=3/2,\ J^{P}=7/2^{-}$, respectively.
\end{abstract}

\date{\today}


\maketitle
\section{Introduction}\label{sec:intro}
In the past decade, with the rapid development of experimental techniques, an increasing number of new hadronic states have been observed~\cite{ParticleDataGroup:2024cfk}.
Exploring the internal structure of these states has become one of the central topics in hadron physics. While many of the observed hadrons can be well accommodated within
the conventional quark model, in which mesons are composed of a quark--antiquark pair~\cite{Godfrey:1985xj} and baryons consist of three quarks~\cite{Capstick:1986ter}, there
also exist states whose internal structures appear to be more intricate. These states, commonly referred to as exotic hadrons, may have internal configurations that include multiquark states, hadronic molecular states, glueballs, quark--gluon hybrid states, or even mixtures of molecular and compact quark components.  All these configurations
are allowed by quantum chromodynamics (QCD), provided that the overall color-singlet condition is satisfied. However, due to quark color confinement and the asymptotic freedom of Quantum Chromodynamics (QCD), the internal structure of hadrons cannot be directly probed experimentally, and first-principles calculations in the low-energy, nonperturbative regime remain highly challenging.  Consequently, theoretical studies of the internal structure of these hadronic states necessarily rely on various phenomenological models and effective theoretical approaches.

Among them, the hadronic molecular picture has proven to be particularly successful. A notable supporting example is the deuteron, which is regarded as a relatively loosely bound
molecular-like state composed of a proton and a neutron. To date, many experimentally observed hadrons can be interpreted as molecular-like states~\cite{Guo:2017jvc}. In particular,
the well-known $X(3872)$, discovered in 2003~\cite{Belle:2003nnu}, is widely considered a $D\bar{D}^{*}$ molecular state~\cite{Brambilla:2019esw,Chen:2022asf,Meng:2022ozq}. Several
subsequently observed hidden-charm pentaquarks, such as the $P_{c}$ and $P_{cs}$ states~\cite{LHCb:2015yax,LHCb:2016ztz,LHCb:2016lve,LHCb:2019kea,LHCb:2020jpq,LHCb:2022ogu}, can also
be interpreted as molecular states, e.g., $D^{(*)}\Sigma_c^{(*)}$ or $D^{(*)}\Xi_c$ configurations~\cite{Chen:2019bip,Guo:2019fdo,Xiao:2019aya,He:2019ify,Xiao:2019mvs,Roca:2015dva,Chen:2015moa,Chen:2015loa,Yang:2015bmv,Huang:2015uda,Du:2019pij}. Moreover, the heavy-quark symmetry~\cite{Isgur:1991wq} suggests that the number of possible molecular states is much larger than those observed experimentally, which has motivated extensive searches for these
yet-to-be-discovered molecular configurations, as discussed in Ref.~\cite{Huang:2024asn}.

In the light-quark sector, there also exist candidates for molecular states, particularly those containing strange quarks.
A prominent example is the $\Lambda(1405)$, which is widely regarded as predominantly a $\bar{K}N$ molecular state~\cite{Oset:1997it}. Similarly, the $K\bar{K}$ interaction can generate
an attractive potential, giving rise to bound molecular states $f_0(980)$ and $a_0(980)$~\cite{Oller:1997ti}. Strong supporting evidence for these molecular interpretations comes from
their natural explanation of the unusually large isospin-breaking effects observed in the $\eta(1405/1475)\to 3\pi$ decay~\cite{Wu:2011yx}. Additionally, the molecular state formed
through the $K\bar{K}^{*}$ interaction can be associated with the experimentally observed $f_1(1285)$~\cite{Roca:2005nm}.
Inspired by the hidden-charm pentaquark molecular states $D\Sigma_c$ and $D^{*}\Sigma_c$, one can replace the $c$ and $\bar{c}$ quarks with $s$ and $\bar{s}$ quarks, leading to the
existence of hidden-strangeness pentaquark nucleon resonances $N(1875)$ and $N(2100)$~\cite{He:2017aps}, whose dominant molecular components are $K\Sigma$ and $K^{*}\Sigma$, respectively.
The antiparticles $\bar{K}$ and $\bar{K}^{*}$ can also interact with the $\Sigma$ baryon to form molecular states, corresponding to the experimentally observed $\Xi(1690)$ and $\Xi(2030)$ resonances~\cite{Sekihara:2015qqa,Khemchandani:2016ftn,Gamermann:2011mq,Miyahara:2016yyh,Hei:2023eqz}. Furthermore, the $\bar{K}$ can interact with the $\Sigma$ and $\Lambda$ baryons
to form molecular states, which can be used to interpret the experimentally observed $\Xi(1620)$ resonance~\cite{Miyahara:2016yyh,Huang:2020taj,Huang:2021ahp}.

Although theoretical calculations indicate that the $\bar{K}^{*}\Sigma^{*}$ interactions can generate $S$-wave molecular states~\cite{Yan:2024usf,Huang:2018ehi,Wang:2023eng,Yang:2022uot,Sarkar:2010saz}, no corresponding states with clear
molecular-structure signatures have been observed experimentally so far.  Indeed, the masses of such molecular states are expected to be located near the thresholds at 2270~MeV (with
$m_{K^{*-}} + m_{\Sigma^{*+}} =2274$~MeV) and 2281~MeV (with $m_{K^{*0}} + m_{\Sigma^{*+}} = 2281$~MeV), respectively.  Experimentally,
a $\Xi(2250)$ state with a mass very close to 2270~MeV has indeed been observed. And one study indicate that this state could be interpreted as a $\Sigma^{*}\bar{K}^{*}$ molecular
state with $J^{P}=3/2^{-}$ or $J^{P}=5/2^{-}$ within the framework of the quark delocalization color screening
model~\cite{Yan:2024usf}.  Moreover, after taking into account the coupled-channel effects, one still find a dynamically generated resonance located very close to the $\Xi(2250)$~\cite{Sarkar:2010saz}.
However, so far there is no experimental evidence indicating that this
state exhibits any exotic features beyond a conventional three-quark
structure~\cite{Aachen-Berlin-CERN-London-Vienna:1969bau,Goldwasser:1970fk,Hassall:1981fs,Jenkins:1983pm,Biagi:1986zj}.

For the $K^{*}\Sigma^{*}$ system, there may exist corresponding molecular state candidates.  Currently, the Particle Data Group (PDG) lists three nucleon
resonances with central masses close to 2281~MeV: the subthreshold states $N(2220)$ and $N(2250)$, and the above-threshold state $N(2300)$. Their spin--parity
quantum numbers are $J^{P}=9/2^{+}$, $J^{P}=9/2^{-}$, and $J^{P}=1/2^{+}$, and they are assigned four, four, and two stars, respectively, by the PDG~\cite{ParticleDataGroup:2024cfk}.  From the quantum numbers corresponding to the particle states, they cannot be identified as the $S$-wave $K^{*}\Sigma^{*}$
molecular states predicted in theoretical studies~\cite{Huang:2018ehi,Wang:2023eng,Yang:2022uot}, and would instead correspond to at least $P$-wave configurations.
By analyzing the available experimental information~\cite{ParticleDataGroup:2024cfk}, either $N(2220)$ or $N(2300)$ might exhibit a molecular-state structure; however,
only $N(2250)$ emerges as a plausible candidate for a $K^{*}\Sigma^{*}$ molecular state, as it has been experimentally observed to decay through the strange channels
$N\eta$ and $K\Lambda$~\cite{Hunt:2018wqz} with branching fractions below $5\%$ and $(1$--$3)\%$, respectively, which can be compared with $(5$--$15)\%$ for the nonstrange $N\pi$ channel~\cite{ParticleDataGroup:2024cfk}.  This suggests that $N(2250)$ may contain an internal $s\bar{s}$ quark component.   For the $K^{*}\Sigma^{*}$ system with isospin $I=3/2$, there also exists a $\Delta(2200)$ with quantum numbers $J^P=7/2^{-}$~\cite{ParticleDataGroup:2024cfk}, which has strange-quark-containing decay channels, $K\Sigma$, whose branching fractions can be compared with the $\pi N$ decay channel. This also suggests that $\Delta(2200)$ could be a plausible candidate for a $K^{*}\Sigma^{*}$ molecular state.

In this study, we investigate whether the interactions of $K^{*}\Sigma^{*}$ can form molecular states, with particular focus on producing high-orbital-angular-momentum states $N(2250)$ and $\Delta(2200)$.  This paper is organized as follows. In Sec.~\ref{Sec: formulism}, we will present the theoretical formalism.  In Sec.~\ref{Sec: results},  the numerical result will be given, followed by discussions and conclusions
in the last section.

\section{FORMALISM AND INGREDIENTS}\label{Sec: formulism}
In this work, we investigate whether the interaction between the $K^{*}$ meson and the $\Sigma^{*}$ baryon can give rise to a hadronic
molecular state corresponding to the $N(2250)$ and $\Delta(2200)$ baryon.  Such molecular states can be obtained by solving the non-relativistic
Schr\"odinger equation
\begin{equation}
\left[-\frac{1}{2\mu} \left( \nabla^2_r - \frac{L(L+1)}{r^2} \right) + V(r) \right] \psi(\vec{r}) = E \, \psi(\vec{r}),
\label{eq:schrodinger}
\end{equation}
where $\mu = (m_{K^{*}} m_{\Sigma^{*}})/(m_{K^{*}} + m_{\Sigma^{*}})$ is the reduced mass of the system, and the radial Laplacian is given by $
\nabla_r^2 = \frac{1}{r^2} \frac{\partial}{\partial r} \left( r^2 \frac{\partial}{\partial r} \right)$.
Here, $L$ denotes the orbital angular momentum of the system, with $L=0$ corresponding to the so-called S-wave molecular state.
The wave function $\psi(\vec{r})$ represents the radial component of the $K^{*}\Sigma^{*}$ molecule, with $E$ denoting its binding energy,
so that the mass of the bound state is given by $m = m_{K^{*}} + m_{\Sigma^{*}} - E$.

\begin{figure}[http]
\begin{center}
\includegraphics[bb=150 640 1050 710, clip, scale=0.75]{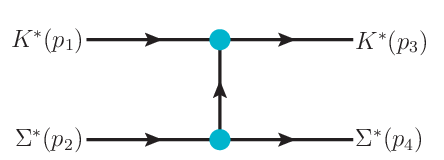}
\caption{Feynman diagram for the process $K^{*}\Sigma^{*} \to K^{*}\Sigma^{*}$, mediated by the exchange of vector mesons ($\rho$, $\omega$) and pseudoscalar
mesons ($\pi$). The four-momenta of the particles are denoted as $p_1$, $p_2$, $p_3$, and $p_4$, respectively.}\label{fig:KstarSigma}
\end{center}
\end{figure}

To obtain the binding energy $E$, the key ingredient is to determine the explicit form of the $K^{*}\Sigma^{*}$ two-body interaction potential $V(r)$.
In the present study, the $K^{*}\Sigma^{*}$  interaction potential is derived within the framework of the one-boson-exchange (OBE) model. According to
the OBE model, the simplest Feynman diagram for the $K^{*}\Sigma^{*} \to K^{*}\Sigma^{*}$ process is the tree-level diagram, as illustrated in
Fig.~\ref{fig:KstarSigma}, where $V$ and $P$ denote the exchanged vector and pseudoscalar mesons, respectively.

In order to compute the potential $V(r)$, the Lagrangians describing the interaction of the three-vector-meson vertex (VVV) and the coupling of two vector mesons
to a pseudoscalar meson (VVP) are given by~\cite{Oset:2002sh,Gonzalez:2008pv}:
\begin{align}
\mathcal{L}_{VVV} &= i g \langle (V_\mu \, \partial_\nu V^\mu - \partial_\nu V_\mu \, V^\mu) V^\nu\rangle , \label{eq:LVVV} \\
\mathcal{L}_{VVP} &= \frac{G'}{\sqrt{2}} \, \epsilon^{\mu\nu\alpha\beta} \, \langle\partial_\mu V_\nu \, \partial_\alpha V_\beta \, P\rangle, \label{eq:LVVP}
\end{align}
where $\epsilon^{\mu\nu\alpha\beta}$ is the Levi-Civit\`a tensor with $\epsilon^{0123} = 1$ and $\langle...\rangle$ denoting a flavor trace.
In Eqs.~(\ref{eq:LVVV}) and~(\ref{eq:LVVP}), $V$ and $P$ denote  the SU(3) matrices collecting the light vector and pseudoscalar meson octets, respectively,
\begin{align}
V_\mu &=
\begin{pmatrix}
\frac{1}{\sqrt{2}}(\rho^0 + \omega) & \rho^+ & K^{*+} \\
\rho^- & -\frac{1}{\sqrt{2}}(\rho^0 - \omega) & K^{*0} \\
K^{*-} & \bar{K}^{*0} & \phi
\end{pmatrix}_\mu, \\
P &=
\begin{pmatrix}
\frac{1}{\sqrt{2}}\pi^0 + \frac{1}{\sqrt{6}}\eta & \pi^+ & K^+ \\
\pi^- & -\frac{1}{\sqrt{2}}\pi^0 + \frac{1}{\sqrt{6}}\eta & K^0 \\
K^- & \bar{K}^0 & -\frac{2}{\sqrt{6}}\eta
\end{pmatrix}.
\end{align}
The coupling constant $G'$ entering the VVP interaction Lagrangian is fixed by the hidden gauge symmetry and is given by~\cite{Oset:2002sh,Gonzalez:2008pv}
\begin{equation}
G' = \frac{3 g'}{4 \pi^2 f},
\end{equation}
with $g' = - (G_V m_\rho)/(\sqrt{2} f^2)$, where $G_V = 55~\mathrm{MeV}$, $m_{\rho}$ is the mass of the $\rho$ meson, and $f = 93~\mathrm{MeV}$ denotes
the pseudoscalar decay constant.

The remaining coupling constant $g$, which governs both the VVV and VPP interactions, can be fixed from experimental data on vector-meson strong decays.
To this end, we consider the effective VPP Lagrangian
\begin{equation}
\mathcal{L}_{VPP} = -i g \langle [P, \partial_\mu P] V^\mu \rangle,
\label{eq:LVPP}
\end{equation}
from which the partial decay width of $K^{*+} \to K^0 \pi^+$ can be derived as
\begin{equation}
\Gamma(K^{*+} \to K^0 \pi^+) = \frac{g^2}{6 \pi m_{K^{*+}}^2} P_{\pi K^*}^3,
\end{equation}
where $P_{\pi K^*}$ is the three-momentum of the $\pi$ in the rest frame of the $K^*$.   Using the experimental value $\Gamma_{K^{*+}} = 50.3 \pm 0.8~\mathrm{MeV}$
together with the physical masses of the involved particles~\cite{ParticleDataGroup:2024cfk}, we obtain g = 4.64.

For the $\Sigma^{*}\Sigma^{*}P$ interaction vertex, we adopt the leading-order effective chiral $SU(3)_L \times SU(3)_R$ Lagrangian~\cite{Holmberg:2018dtv,Lutz:2001yb,Copeland:2020ljp,Pascalutsa:2006up},
\begin{align}
\mathcal{L}_{TTP}= -\frac{H_A}{2}\,\bar{T}^{\mu}_{abc}\,\gamma_{\nu}\gamma_5\,(u^{\nu})^{c}_{\ d}\,T^{abd}_{\mu},
\end{align}
where $a, b, c$ denote flavor indices, and $u_\mu$ is the axial-vector combination of the pseudoscalar-meson
fields and their derivatives,
\begin{equation}
u_\mu = i \left( u^\dagger \partial_\mu u - u \, \partial_\mu u^\dagger \right),
\label{eq:axialvector}
\end{equation}
with $u^2 = U = \exp {(iP/f)}$.
At leading order, one has $u^{\nu} = -\sqrt{2}\, \partial^{\nu} P / f$.
The coupling constant $H_A$ is not uniquely determined due to the lack of direct experimental observables.
Theoretically, it can be estimated using
\textit{large-$N_c$} methods, yielding two different results: $H_A = 9g_A/5 \approx 2.27$~\cite{Pascalutsa:2006up,Ledwig:2011cx}
and $H_A = 9 F - 3 D \approx 1.74$~\cite{Semke:2005sn,Dashen:1993as}.  In this work, we will perform calculations using both of these values separately.
$T^\mu_{abc}$ represents the baryon decuplet field, described by spin-$3/2$ Rarita--Schwinger
fields and expressed as a fully symmetric tensor in flavor space.
It contains the $\Delta$ isobar, the $\Sigma^{*}$, $\Xi^{*}$, and the triply-strange
$\Omega^{-}$ states, whose explicit flavor representation is given by~\cite{Holmberg:2018dtv,Lutz:2001yb,Copeland:2020ljp,Pascalutsa:2006up}
\begin{widetext}
\begin{align}
T_\mu =
\left\{
\begin{pmatrix}
\Delta^{++} & \dfrac{1}{\sqrt{3}}\Delta^{+} & \dfrac{1}{\sqrt{3}}\Sigma^{*+} \\
\dfrac{1}{\sqrt{3}}\Delta^{+} & \dfrac{1}{\sqrt{3}}\Delta^{0} & \dfrac{1}{\sqrt{6}}\Sigma^{*0} \\
\dfrac{1}{\sqrt{3}}\Sigma^{*+} & \dfrac{1}{\sqrt{6}}\Sigma^{*0} & \dfrac{1}{\sqrt{3}}\Xi^{*0}
\end{pmatrix},
\begin{pmatrix}
\dfrac{1}{\sqrt{3}}\Delta^{+} & \dfrac{1}{\sqrt{3}}\Delta^{0} & \dfrac{1}{\sqrt{6}}\Sigma^{*0} \\
\dfrac{1}{\sqrt{3}}\Delta^{0} & \Delta^{-} & \dfrac{1}{\sqrt{3}}\Sigma^{*-} \\
\dfrac{1}{\sqrt{6}}\Sigma^{*0} & \dfrac{1}{\sqrt{3}}\Sigma^{*-} & \dfrac{1}{\sqrt{3}}\Xi^{*-}
\end{pmatrix},
\begin{pmatrix}
\dfrac{1}{\sqrt{3}}\Sigma^{*+} & \dfrac{1}{\sqrt{6}}\Sigma^{*0} & \dfrac{1}{\sqrt{3}}\Xi^{*0} \\
\dfrac{1}{\sqrt{6}}\Sigma^{*0} & \dfrac{1}{\sqrt{3}}\Sigma^{*-} & \dfrac{1}{\sqrt{3}}\Xi^{*-} \\
\dfrac{1}{\sqrt{3}}\Xi^{*0} & \dfrac{1}{\sqrt{3}}\Xi^{*-} & \Omega^{-}
\end{pmatrix}
\right\}.\nonumber
\end{align}
\end{widetext}

To evaluate the diagrams shown in Fig.~\ref{fig:KstarSigma}, it is necessary to specify
the effective Lagrangian densities that describe the interaction vertices
between vector mesons and the $\Sigma^*$ baryon. These interaction terms are given by~\cite{He:2017aps,Matsuyama:2006rp}:
\begin{align}
\mathcal{L}_{\Sigma^* \Sigma^* \rho} &=- g_{\Sigma^* \Sigma^* \rho} \, \bar{\Sigma}^{*}_\mu
\Big( \gamma_\nu - \frac{\kappa_{\Sigma^* \Sigma^* \rho}}{2 m_{\Sigma^*}} \sigma_{\nu \alpha} \partial^\alpha \Big)
\rho^\nu \cdot T \, \Sigma^*_\mu, \label{eq:L_SigStar_rho} \\
\mathcal{L}_{\Sigma^* \Sigma^* \omega} &=- g_{\Sigma^* \Sigma^* \omega} \, \bar{\Sigma}^{*}_\mu
\Big( \gamma_\nu - \frac{\kappa_{\Sigma^* \Sigma^* \omega}}{2 m_{\Sigma^*}} \sigma_{\nu \alpha} \partial^\alpha \Big)
\omega^\nu \, \Sigma^*_\mu, \label{eq:L_SigStar_omega} \\
\mathcal{L}_{\Sigma^* \Sigma^* \phi} &=- g_{\Sigma^* \Sigma^* \phi} \, \bar{\Sigma}^{*}_\mu
\Big( \gamma_\nu - \frac{\kappa_{\Sigma^* \Sigma^* \phi}}{2 m_{\Sigma^*}} \sigma_{\nu \alpha} \partial^\alpha \Big)
\phi^\nu \, \Sigma^*_\mu, \label{eq:L_SigStar_phi}
\end{align}
where $m_{\Sigma^{*}}$ is the mass of the $\Sigma^{*}$ baryon and $\sigma^{\mu\nu}=\frac{i}{2}(\gamma^{\mu}\gamma^{\nu}-\gamma^{\nu}\gamma^{\mu})$.
The coupling constants are given by $g_{\Sigma^* \Sigma^* \rho} = g_{\Delta\Delta\rho}$, $g_{\Sigma^* \Sigma^* \omega} = - g_{\Delta\Delta\rho}$, and
$g_{\Sigma^* \Sigma^* \phi} = g_{\Delta\Delta\rho}/\sqrt{2}$, with $g_{\Delta\Delta\rho} = 6.1994$. These values are fixed by SU(3) flavor symmetry,
and a detailed derivation can be found in Ref.~\cite{Matsuyama:2006rp}. Similarly, the tensor coupling constants $\kappa_{\Sigma^*\Sigma^* V}$
($V = \rho, \omega, \phi$) follow the same SU(3) relations as the corresponding vector couplings:
$\kappa_{\Sigma^*\Sigma^*\rho} = - \kappa_{\Sigma^*\Sigma^*\omega} = \sqrt{2}\, \kappa_{\Sigma^*\Sigma^*\phi} = \kappa_{\Delta\Delta\rho} = 6.1.$
The isospin operator $T$ for the $\Sigma^*$ baryon is represented in the $I=1$ isospin space by the matrices
\begin{align}
T_z &=
\begin{pmatrix}
1 & 0 & 0  \\
0 & 0 & 0  \\
0 & 0 & -1
\end{pmatrix},\qquad
T_x =
\begin{pmatrix}
0 & \sqrt{\frac{1}{2}} & 0  \\
\sqrt{\frac{1}{2}} & 0 & \sqrt{\frac{1}{2}}  \\
0 & \sqrt{\frac{1}{2}} & 0
\end{pmatrix},\nonumber\\
T_y &=
\begin{pmatrix}
0 & -\frac{i}{\sqrt{2}} & 0  \\
\frac{i}{\sqrt{2}} & 0 & -\frac{i}{\sqrt{2}}  \\
0 & \frac{i}{\sqrt{2}} & 0
\end{pmatrix},
\end{align}
where the basis states are ordered as $|I_3 = 1\rangle$, $|0\rangle$, and $|-1\rangle$.
These matrices satisfy the SU(2) algebra, $[T_a, T_b] = i \varepsilon_{abc} T_c$,
and act on the three-component isospin multiplet of the $\Sigma^*$ resonance.

With the ingredients introduced above, we are now in a position to write down the general expressions for the scattering amplitudes corresponding to the Feynman diagrams shown in Fig.~\ref{fig:KstarSigma}.
The contributions from vector-meson exchange can be written as
\begin{align}
{\cal{M}}_{\cal{V}}&=\sum_{V=\rho^0,\omega,\phi}-{\cal{I}}_{V}\frac{igg_{\Sigma^{*+}\Sigma^{*+}V}}{\sqrt{2}}\bar{u}_{\alpha}(s_4,p_4)[\gamma_{\beta}+\frac{\kappa_{\Sigma^{*}\Sigma^{*}V}}{4m_{\Sigma^{*+}}}(\gamma_{\beta}q\!\!\!/-q\!\!\!/\gamma_{\beta})]\nonumber\\
                  &\times{}u^{\alpha}(s_2,p_2)\frac{-g^{\beta\eta}+q^{\beta}q^{\eta}/m^2_{V}}{q^2-m^2_{V}}[-p_3\cdot\epsilon_1\epsilon^{\dagger}_{3\eta}-p_1\cdot\epsilon^{\dagger}_3\epsilon_{1\eta}\nonumber\\
                  &+\epsilon^{\dagger}_3\cdot{}\epsilon_1(p_{1}+p_3)_{\eta}+q\cdot{}\epsilon^{\dagger}_3\epsilon_{1\eta}-q\cdot{}\epsilon_1\epsilon^{\dagger}_{3\eta}],\label{eq16}\\
{\cal{M}}_{\pi}&=\frac{H_AG^{'}}{6f}\bar{u}_{\mu}(s_4,p_4)q\!\!\!/\gamma_5u_{\mu}(s_2,p_2)\frac{1}{q^2-m^2_{\pi}}\epsilon^{\nu\eta\alpha\beta}\nonumber\\
               &\times{}p_{3\alpha}p_{1\nu}\epsilon_{3\beta}^{\dagger}\epsilon_{1\eta},
\end{align}
where $q = p_3 - p_1 = p_2 - p_4$, and the coefficients $\mathcal{I}_{\rho^0}=-1$, $\mathcal{I}_{\omega}=1$, and $\mathcal{I}_{\phi}=-\sqrt{2}$ correspond to the $\rho^0$, $\omega$, and $\phi$ meson exchanges, respectively.
The polarization vector $\epsilon_\mu(s,p)$, labeled by the spin projection $s$ and four-momentum $p$, describes the wave function of a spin-one particle and can be expressed as
\begin{align}
\epsilon_\mu(s=0) &=
\begin{pmatrix}
0 \\ 0 \\ 0 \\ 1
\end{pmatrix}; \qquad
\epsilon_\mu(s=\pm) =
\frac{1}{\sqrt{2}}
\begin{pmatrix}
0 \\ \mp 1 \\ -i \\ 0
\end{pmatrix}.
\end{align}
The spinors $u_{\alpha}(s_1,p_1)$ and $\bar{u}_{\alpha}(s_3,p_3)$ in Eq.~\ref{eq16} denote the spin wave functions of the initial and final $\Sigma^{*}$ baryons, respectively, with
$\bar{u}_{\alpha}(s_3,p_3) = u^{\dagger}_{\alpha}(s_3,p_3)\, \gamma^0$.
In practical calculations, $u_{\alpha}(s_1,p_1)$ is often expressed as a state obtained by coupling a spin-$1/2$ particle wave function with that of a spin-$1$ particle, i.e.,
\begin{align}
u_{\alpha}(s,p)       &= \sum_{s_{1/2}, s_1} C^{3/2,s}_{1/2,s_{1/2};1,s_1} u(s_{1/2}, p)\epsilon_{\alpha}(s_1, p),\label{eq17}\\
\bar{u}_{\alpha}(s,p) &= u^{\dagger}_{\alpha}(s,p)\gamma^0,\nonumber
\end{align}
where $s$, $s_{1/2}$, and $s_1$ denote the third components of the spin, and $C^{3/2,s}_{1/2,s_{1/2};1,s_1}$ is the Clebsch--Gordan coefficient~\cite{ParticleDataGroup:2024cfk}.

For the spin-$1/2$ Dirac spinor $u(s_{1/2},\vec{q})$ and $\bar{u}(s_{1/2},\vec{q})$, we adopt the standard representation
\begin{align}
u(s_{1/2}, \vec{q}) &= \sqrt{\frac{E + m}{2 m}}
\begin{pmatrix}
1 \\[1mm]
\frac{\vec{\sigma} \cdot \vec{q}}{E + m}
\end{pmatrix} \chi_{s_{1/2}}, \label{eq:dirac1}\\
\bar{u}(s_{1/2}, \vec{q}) &= u^\dagger(\vec{q}, s) \gamma^0
= \chi_{s_{1/2}}^\dagger \sqrt{\frac{E + m}{2 m}}
\begin{pmatrix}
1 & - \frac{\vec{\sigma} \cdot \vec{q}}{E + m}
\end{pmatrix}, \label{eq:dirac2}
\end{align}
where $\chi_{1/2} = \begin{pmatrix} 1 \\ 0 \end{pmatrix}$ and $\chi_{-1/2} = \begin{pmatrix} 0 \\ 1 \end{pmatrix}$.
Here, $E$, $\vec{q}$, and $m$ denote the energy, three-momentum, and mass of the baryon, satisfying the on-shell condition $E^2 = \vec{q}^{\,2} + m^2$.

To simplify the amplitude, we adopt the following kinematic setup.
The initial-state four-momenta are chosen as
\begin{equation}
p_1 = (E_1,\vec{p}), \qquad p_2 = (E_2,-\vec{p}),
\end{equation}
while the final-state four-momenta are
\begin{equation}
p_3 = (E_1,\vec{p}\,'), \qquad p_4 = (E_2,-\vec{p}\,').
\end{equation}
With these definitions, the four-momentum transferred by the exchanged particle reads
\begin{equation}
q = p_3 - p_1 = p_2 - p_4 = (0,\vec{p}\,'-\vec{p}) \equiv (0,\vec{q}).
\end{equation}
By adopting the nonrelativistic approximation and retaining terms up to order $1/m^2$, the scattering amplitudes are
\begin{align}
{\cal{M}}_{\cal{V}}&=\sum_{V=\rho^0,\omega,\phi}{\cal{I}}_{V}\frac{igg_{\Sigma^{*+}\Sigma^{*+}V}}{\sqrt{2}}(\vec{\epsilon}_3^{\dagger}\cdot{}\vec{\epsilon}_12m_{K^{*}}{\cal{X}}\nonumber\\
                   &-4{\cal{Y}}\cdot{}{\cal{A}})\frac{1}{\vec{q}^2+m^2_V}(\frac{\Lambda^2-m^2}{\Lambda^2-q^2})^2,\\
{\cal{M}}_{\pi}&=\frac{i2m_{K^{*}}H_AG^{'}}{3f}\frac{i(\vec{\sigma}_2\cdot\vec{q})[\vec{q}\cdot{}(\vec{\epsilon}_1\times\vec{\epsilon}^{\dagger}_3)]}{\vec{q}^2+m^2_{\pi}}\nonumber\\
               &\times(\frac{\Lambda^2-m^2}{\Lambda^2-q^2})^2,
\end{align}
with
\begin{align}
{\cal{X}}&=1+\frac{\vec{k}^2}{4m^2_{\Sigma^{*}}}-\frac{\vec{q}^2(1+4\kappa_{\Sigma^{*}\Sigma^{*}V})}{16m^2_{\Sigma^{*}}}-i\frac{\vec{\sigma}_2\cdot(\vec{k}\times\vec{q})}{4m^2_{\Sigma^{*}}}\nonumber\\
{\cal{Y}}&=2i(\vec{\sigma}_2\times\vec{q})+\frac{1}{4m^2_{\Sigma^*}}[2(\vec{q}\times\vec{k})~\vec{\sigma}_2\cdot(\vec{k}-\frac{1}{2}\vec{q})\nonumber\\
         &+i(\vec{q}\cdot{}\vec{k}\vec{q}-\vec{q}^2\vec{k})+(\vec{k}^2-\frac{1}{4}\vec{q}^2)(\vec{\sigma}_2\times\vec{q})],\nonumber\\
{\cal{A}}&=(\frac{3}{2}\vec{q}+\vec{k})\cdot\vec{\epsilon}_1\vec{\epsilon}^{\dagger}_3-(\frac{3}{2}\vec{q}-\vec{k})\cdot\vec{\epsilon}^{\dagger}_3\vec{\epsilon}_1-2\vec{k}\vec{\epsilon}^{\dagger}_3\cdot{}\vec{\epsilon}_1.\nonumber
\end{align}
In the above expression, $\vec{k}=(\vec{p}+\vec{p}^{'})/2$, and the factor $(\Lambda^2 - m_i^2)^2/(\Lambda^2 - q_i^2)^2$ is introduced as a form factor to account for the finite-size effects of the exchanged meson, where $m_i$ and $q_i$ represent the mass and four-momentum of the $i$-th exchanged meson, respectively.
The cutoff parameter $\Lambda_i$ is taken as $\Lambda_i = m_i + \alpha \, \Lambda_{\rm QCD}$, with $\Lambda_{\rm QCD} = 220~\text{MeV}$, where $\alpha$ is treated as a free parameter and will be discussed later. This form factor suppresses contributions from high-momentum transfer, reflecting the fact that the meson is not a point particle but possesses an internal structure.

Once the scattering amplitude is obtained, the interaction potential in momentum space can be derived directly using the Breit approximation,
\begin{align}
{\cal V}(\vec{q})
= -\frac{{\cal M}}{\sqrt{\prod_{i} 2m_i \, \prod_{f} 2m_f}},
\end{align}
where $m_i$ and $m_f$ denote the masses of the initial and final states, respectively.
By performing a Fourier transformation of the momentum--space potential, one obtains
the coordinate--space potential $V(\vec{r})$ entering Eq.~\ref{eq:schrodinger}.
The required Fourier-transform relations are
\begin{align}
&{\cal F}\!\left\{\frac{1}{\vec{q}^{\,2}+m^{2}}
    \left(\frac{\Lambda^{2}-m^{2}}{\Lambda^{2}+\vec{q}^{\,2}}\right)^2
\right\}= Y_{1}(\Lambda,m,r),\nonumber\\[2mm]
&{\cal F}\!\left\{
    \frac{\vec{q}^{\,2}}{\vec{q}^{\,2}+m^{2}}
    \left(\frac{\Lambda^{2}-m^{2}}{\Lambda^{2}+\vec{q}^{\,2}}\right)^{2}
\right\}
= -\nabla_{r}^{2} Y_{1}(\Lambda,m,r), \nonumber\\[2mm]
&{\cal F}\!\left\{
    \frac{\vec{k}^{\,2}}{\vec{q}^{\,2}+m^{2}}
    \left(\frac{\Lambda^{2}-m^{2}}{\Lambda^{2}+\vec{q}^{\,2}}\right)^{2}
\right\}
= \frac{1}{4}\nabla_{r}^{2} Y_{1}(m,r)
  - \frac{1}{2}\{\nabla_{r}^{2},Y_{1}(m,r)\}, \nonumber\\[2mm]
&{\cal F}\!\left\{
    \frac{i\,\vec{\sigma}\!\cdot\!(\vec{q}\times\vec{k})}
         {\vec{q}^{\,2}+m^{2}}
    \left(\frac{\Lambda^{2}-m^{2}}{\Lambda^{2}+\vec{q}^{\,2}}\right)^{2}
\right\}
= \vec{\sigma}\cdot\vec{L}\,
  \frac{1}{r}\frac{\partial Y_{1}(m,r)}{\partial r}, \nonumber\\[2mm]
&{\cal F}\!\left\{
    \frac{(\vec{A}\cdot\vec{q})(\vec{B}\cdot\vec{q})}
         {\vec{q}^{\,2}+m^{2}}
    \left(\frac{\Lambda^{2}-m^{2}}{\Lambda^{2}+\vec{q}^{\,2}}\right)^{2}
\right\}
= \frac{1}{3}(\vec{A}\!\cdot\!\vec{B})
  \big(-\nabla_{r}^{2}Y_{1}(\Lambda,m,r)\big) \nonumber\\
&\hspace{3cm}
  + \frac{1}{3} S(\hat{r},\vec{A},\vec{B})
    \left(
      - r\frac{\partial}{\partial r}
        \frac{1}{r}\frac{\partial}{\partial r}
        Y_{1}(\Lambda,m,r)
    \right), \nonumber\\[2mm]
&{\cal F}\!\left\{
    \frac{(\vec{A}\cdot\vec{k})(\vec{B}\cdot\vec{q})}
         {\vec{q}^{\,2}+m^{2}}
    \left(\frac{\Lambda^{2}-m^{2}}{\Lambda^{2}+\vec{q}^{\,2}}\right)^{2}
\right\}
= \frac{1}{3}(\vec{A}\!\cdot\!\vec{B})\,\nabla_{r}^{2}Y_{1}(\Lambda,m,r)
  \nonumber\\
&\hspace{3cm}
  + \frac{1}{3}\!
    \left(
      r\frac{\partial}{\partial r}
      \frac{1}{r}\frac{\partial}{\partial r}
      Y_{1}(\Lambda,m,r)
    \right)
  - \frac{2}{3}
    \frac{\partial Y_{1}(\Lambda,m,r)}{\partial r}\nonumber\\
&\hspace{3cm}\times    \left(S(\hat{r},\vec{A},\vec{B})+\vec{A}\!\cdot\!\vec{B}\right)
    \frac{\partial}{\partial r}.
\end{align}
Here ${\cal F}$ denotes the Fourier transformation, and
$\nabla_r^{2} = \dfrac{1}{r^{2}} \dfrac{\partial}{\partial r}
\!\left( r^{2} \dfrac{\partial}{\partial r} \right)$
is the radial Laplacian.
The operator $\vec{\sigma}\!\cdot\!\vec{L}$ represents the spin--orbit coupling, while
$S(\hat{r},\vec{x},\vec{y}) = 3(\hat{r}\!\cdot\!\vec{x})(\hat{r}\!\cdot\!\vec{y}) - \vec{x}\!\cdot\!\vec{y}$
denotes the tensor operator.  The function $Y_1(\Lambda,m,r)$ and the recoil correction $F_1$ are defined as
\begin{align}
&Y_1(\Lambda,m,r) = \frac{1}{4\pi r} ( e^{-m r} - e^{-\Lambda r} )- \frac{\Lambda^2 - m^2}{8 \pi \Lambda} e^{-\Lambda r}, \nonumber\\
&F_1(m) = \{ \nabla_r^2, Y_1(\Lambda,m,r) \}= \nabla_r^2 Y_1(\Lambda,m,r) + Y_1(\Lambda,m,r) \nabla_r^2.\nonumber
\end{align}

In the above Fourier-transform relations and recoil terms, the operators $\frac{\partial}{\partial r}$ and $\nabla_r^2$ act explicitly on the radial wave function $\psi(\vec{r})$ appearing in Eq.~\ref{eq:schrodinger}, which is expanded using the Gaussian expansion method~\cite{Hiyama:2003cu}.
To evaluate the operators $S(\hat{r},\vec{A},\vec{B})$, $\vec{\sigma}\cdot\vec{L}$, and related terms, the system's spin--orbit wave function is required. This wave function can be expressed as
\begin{align}
|K^{*}\Sigma^{*}(^{2S+1}L_J)\rangle= \sum^{m,m^{'}}_{m_S,m_L}C^{S,m_S}_{3/2,m;1,m'}C^{J,M}_{S,m_S;L,m_L}{\cal X}_{3/2,m}\epsilon_{1,m'}|Y_{L,m_L}\rangle,\nonumber
\end{align}
where $\chi_{3/2,3/2}=\chi_{1/2}\epsilon_{1,1}$, $\chi_{3/2,1/2}=\sqrt{\frac{2}{3}}\chi_{1/2}\epsilon_{1,0}+\sqrt{\frac{1}{3}}\chi_{-1/2}\epsilon_{1,1}$, $\chi_{3/2,-1/2}=\sqrt{\frac{1}{3}}\chi_{1/2}\epsilon_{1,-1}+\sqrt{\frac{2}{3}}\chi_{-1/2}\epsilon_{1,0}$, and $\chi_{3/2,-3/2}=\chi_{-1/2}\epsilon_{1,-1}$.
$|Y_{L,m_L}\rangle$ denotes the spherical harmonics function, and $M$, $m_S$, and $m_L$ are the $z$-components of the total ($J$), spin ($S$), and orbital ($L$)
angular momenta, respectively. Here, $m$ and $m'$ are the $z$-components of spin $3/2$ and $1$, respectively.   The possible spin states of the system are
listed in Table.~\ref{tab:spins}, and the computed values of $S(\hat{r},\vec{A},\vec{B})$, $\vec{\sigma}\cdot\vec{L}$, and related operators are presented
in Table.~\ref{tableqw} and Table.~\ref{tableqw-1}.  It should be noted that Table.~\ref{tab:spins} includes data only up to $F$-waves.
\begin{table}[htbp]
\caption{Possible quantum numbers for $K^{*}\Sigma^{*}$ systems involved in our calculation.
The first column contains the spin-parity quantum numbers corresponding to the channels.
$A \sim B$ stands for the mixing effect between $A$ and $B$.}\label{tab:spins}
\centering
\setlength{\tabcolsep}{16pt}
\begin{tabular}{c c }
\hline\hline
$J^P$ & $K^{*}\Sigma^{*}$ \\
\hline
~~~$1/2^-$ & $|{}^2S_{1/2} \sim {}^4 D_{1/2} \sim {}^6 D_{1/2}\rangle$  \\
~~~$1/2^{+}$ & $|{}^2P_{1/2} \sim {}^4P_{1/2} \sim {}^6F_{1/2}\rangle$ \\
~~~$3/2^-$ & $|{}^4S_{3/2} \sim {}^2 D_{3/2} \sim {}^4 D_{3/2}\sim {}^6 D_{3/2}\rangle$\\
~~~$3/2^{+}$ & $|{}^2P_{3/2} \sim {}^4P_{3/2} \sim {}^6P_{3/2} \sim {}^4F_{3/2} \sim {}^6F_{3/2}\rangle$ \\
~~~$5/2^-$ & $|{}^6S_{5/2} \sim {}^2 D_{5/2} \sim {}^4 D_{5/2}\sim {}^6 D_{5/2}\rangle$ \\
~~~$5/2^+$ & $|{}^4P_{5/2} \sim {}^6 P_{5/2} \sim {}^2 F_{5/2}\sim {}^4 F_{5/2}\sim {}^6 F_{5/2}\rangle$ \\
~~~$7/2^-$ & $|{}^4 D_{7/2}\sim {}^6 D_{7/2}\rangle$ \\
~~~$7/2^+$ & $|{}^6 P_{7/2} \sim {}^4 F_{7/2}\sim {}^6 F_{7/2}\rangle$ \\
~~~$9/2^-$ & $|{}^6 D_{9/2}\rangle$ \\
~~~$9/2^+$ & $| {}^4 F_{9/2}\sim {}^6 F_{9/2}\rangle$ \\
~~~$11/2^+$ & $| {}^6 F_{11/2}\rangle$ \\
\hline\hline
\end{tabular}
\end{table}

\begin{table*}[t!]
	\centering
	\caption{The matrix elements of two-body interaction operators for $VB$ systems.}\label{tableqw}
     \setlength{\tabcolsep}{1.0pt}
	\begin{tabular}{cccccccc}
		\hline\hline
		$VB\to{}VB$    &$1/2^{-}$                    & $3/2^{-}$                               &$3/2^{+}$                  &$7/2^{-}$         \\
${\cal{Z}}$		&($^{2}S_{J}$,$^{4}D_{J}$, $^{6}D_{J}$)& ($^{4}S_{J},{}^{2}D_{J},{}^{4}D_{J},{}^{6}D_{J}$) &($^2P_{J},{}^{4}P_{J}, {}^{6}P_{J},{}^{4}F_{J}, {}^{6}F_{J}$)
&($^4D_{J}, ^{6}D_J$)   \\
	$\vec{\sigma}\cdot{}\vec{L}$&
		$\left(
		\begin{array}{ccc}
			0        &   0    & 0   \\
			0       &   -9    & 1  \\
            0       &   1     & -14   \\
		\end{array}
		\right)$  &
		$\left(
		\begin{array}{cccc}
			0        &  0         &   0   & 0   \\
			0        &  -3        &   3/\sqrt{2}   &0  \\
			0        &  3/\sqrt{2}          &   -6   & \sqrt{7/2}  \\
            0        &  0         &   \sqrt{7/2}   &-11  \\
		\end{array}
		\right)$ &
		$\left(
		\begin{array}{ccccc}
			1    &   \sqrt{5/2}  & 0  & 0    &0   \\
			\sqrt{5/2}    &   -2  & \sqrt{3/2}  & 0   &0   \\
            0    &   \sqrt{3/2}  & -7 & 0    &0  \\
            0    &   0  & 0  & -12  &2    \\
             0    &   0  & 0  & 2  &-17    \\
		\end{array}
		\right)$&
		$\left(
		\begin{array}{cc}
			6        &   \sqrt{6}       \\
			\sqrt{6}       &   1       \\
		\end{array}
		\right)$
		\\	
$i\vec{\sigma}\cdot(\vec{\epsilon}_1\times\epsilon_3^{\dagger})$
		&$\left(
		\begin{array}{ccc}
			\frac{-5}{9}                 &   0                      & 0   \\
			0                             &   \frac{-2}{9}           & 0      \\
            0                             &   0                      & \frac{1}{3}        \\
		\end{array}
		\right)$
		& $\left(
			\begin{array}{cccc}
			    \frac{-26}{45}                      & 0                          &   0                          & 0                     \\
				0                                   & \frac{-5}{9}               &   \frac{7\sqrt{2/5}}{45}      &\frac{2\sqrt{7/5}}{15}  \\
				0                                   & \frac{7\sqrt{2/5}}{45}      &   \frac{-2}{225}             &\frac{-\sqrt{14}}{75}  \\
                0                                   &  \frac{2\sqrt{7/5}}{15}     &   \frac{-\sqrt{14}}{75}     &\frac{29}{75}          \\
			\end{array}
			\right)$
		& $\left(
			\begin{array}{ccccc}
			\frac{-5}{9}                      & \frac{-7\sqrt{2}}{45}           &\frac{2}{15\sqrt{3}}              &0                       & 0                    \\
			\frac{-7\sqrt{2}}{45}              & \frac{-22}{75}                 &\frac{-7\sqrt{2/3}}{75}            &0                       & 0                    \\
            \frac{2}{15\sqrt{3}}               & \frac{-7\sqrt{2/3}}{75}        & \frac{-1}{25}                   &0                       & 0                    \\
            0                                  &  0                             & 0                               & \frac{-22}{75}         & \frac{4}{75}         \\
            0                                  &  0                             & 0                               & \frac{4}{75}           & \frac{47}{75}\\
			\end{array}
			\right)$
		&$\left(\begin{array}{cc}
				\frac{-10}{21}                    & \frac{-2\sqrt{2/3}}{21}   \\
			    \frac{-2\sqrt{2/3}}{21}           &  \frac{1}{9}           \\
			\end{array}\right) $\\
$\vec{\epsilon}_{1}\cdot\vec{\epsilon}^{\dagger}_{3}$	
           &$\left(
		\begin{array}{ccc}
			1        &   0    & 0   \\
			0       &   1    & 0  \\
            0       &   0     & 1   \\
		\end{array}
		\right)$
			& $\left(
			\begin{array}{cccc}
			1 &  0   &   0  & 0  \\
				0 &  1 &   0  &0  \\
				0  &  0  &   1  &0  \\
                0 &0&0&1\\
			\end{array}
			\right)$
			& $\left(
			\begin{array}{ccccc}
			1          & 0    &0 & 0& 0 \\
			0  & 1     &0   &0  & 0\\
            0  &  0  & 1  & 0 & 0\\
            0  &  0  & 0  & 1 & 0\\
            0  &  0  & 0  & 0 & 1\\
			\end{array}
			\right)$
			&$\left(
			\begin{array}{cc}
				1        & 0  \\
				0  &  1          \\
			\end{array}
			\right)$ \\
$iS(\hat{r},\vec{\epsilon}_3^{\dagger}\times\vec{\sigma},\vec{\epsilon}_1)$	
             &$\left(
		\begin{array}{ccc}
			0                            &   \frac{-13}{45\sqrt{5}}           & \frac{-7}{15\sqrt{5}}   \\
			\frac{-22}{45\sqrt{5}}        &   \frac{8}{45}                   & \frac{-2}{15}         \\
            \frac{17}{15\sqrt{5}}         &   \frac{1}{15}                   & \frac{4}{15}          \\
		\end{array}
		\right)$
            & $\left(
			\begin{array}{cccc}
			    0                      & \frac{19\sqrt{2/5}}{45}     &   \frac{-44}{225}             & \frac{-19\sqrt{2/7}}{75}    \\
			 \frac{22\sqrt{2/5}}{45}    & \frac{1}{9}                &   \frac{43\sqrt{2/5}}{315}     &\frac{12}{35\sqrt{35}}    \\
				\frac{-56}{225}        & \frac{-2\sqrt{10}}{63}    &   \frac{16}{225}             &\frac{-178\sqrt{2/7}}{525}  \\
                \frac{-16\sqrt{2/7}}{75} & \frac{-8}{7\sqrt{35}}     &   \frac{227\sqrt{2/7}}{525}   &\frac{121}{525}           \\
			\end{array}
			\right)$
           & $\left(
			\begin{array}{ccccc}
			\frac{1}{9}                       &\frac{11\sqrt{2}}{75}             &\frac{32}{75\sqrt{3}}          &\frac{-12\sqrt{2}}{175}   & \frac{-23\sqrt{2}}{525} \\
			\frac{2\sqrt{2}}{75}             & \frac{-88}{1125}                  &\frac{-166\sqrt{2/3}}{375}     &\frac{-88}{875}           & \frac{-232}{2625}        \\
            \frac{-76}{75\sqrt{3}}           & \frac{59\sqrt{2/3}}{375}          & \frac{17}{125}                & \frac{34\sqrt{6}}{875}   & \frac{-33\sqrt{6}}{875}     \\
            \frac{-97\sqrt{2}}{525}          & \frac{-884}{2625}                 & \frac{29\sqrt{6}}{875}        & \frac{56}{375}           & \frac{-104}{2625}        \\
            \frac{79\sqrt{2}}{525}           & \frac{8}{2625}                    & \frac{-23\sqrt{6}}{875}       & \frac{496}{2625}         & \frac{89}{375}        \\
			\end{array}
			\right)$
        &$\left(\begin{array}{cc}
				\frac{44}{735}                  & \frac{52\sqrt{2/3}}{735}     \\
			    \frac{166\sqrt{2/3}}{735}       &  \frac{-227}{2205}           \\
			\end{array}\right) $\\
		\hline
   $VB\to{}VB$    &$7/2^{+}$                    & $5/2^{-}$                               &$5/2^{+}$                   &$9/2^{+}$                   \\
${\cal{Z}}$		&($^{6}P_{J},^{4}F_{J},^{6}F_{J}$)& ($^{6}S_{J},^{2}D_{J},^{4}D_{J},^{6}D_{J}$) &($^{4}P_J, ^{6}P_{J}, ^{2}F_{J}, ^{4}F_{J}, ^{6}F_J$) &($^{4}F_{J},^{6}F_{J}$) \\
     $\vec{\sigma}\cdot{}\vec{L}$&
		$\left(
		\begin{array}{ccc}
			5       &   0    & 0   \\
			0       &   0    & 5/\sqrt{2}  \\
            0       &   5/\sqrt{2}     & -5   \\
		\end{array}
		\right)$ &
		$\left(
		\begin{array}{cccc}
			0       &   0    & 0   & 0\\
			0       &   2    & \sqrt{7} &0 \\
            0       &   \sqrt{7}    & -1  & \sqrt{6}\\
            0       &     0   & \sqrt{6}  & -6\\
		\end{array}
		\right)$&
		$\left(
		\begin{array}{ccccc}
			3                    &   \sqrt{7/3}           & 0                        &  0                    &  0 \\
		\sqrt{7/3}       &   -2                           & 0                        &  0                    &  0 \\
            0                    &   0                            & -4                       & 3\sqrt{6}/2   &  0  \\
            0                    &     0                          & 3\sqrt{6}/2      & -7                    & 5\sqrt{2}/2    \\
            0                    &     0                          & 0                        & 5\sqrt{2}/2   & -12   \\
		\end{array}
		\right)$&
		$\left(
		\begin{array}{ccccc}
			9                    &   \sqrt{14}           \\
		\sqrt{14}                &   4                   \\
		\end{array}
		\right)$\\
 $i\vec{\sigma}\cdot(\vec{\epsilon}_1\times\epsilon_3^{\dagger})$
       & $\left(
		\begin{array}{ccc}
			\frac{-1}{7}                         &   0                                  & 0   \\
			0                                     &   \frac{-26}{189}                    & \frac{-\sqrt{2}}{21}      \\
            0                                     &   \frac{-\sqrt{2}}{21}              & \frac{25}{63}        \\
		\end{array}
		\right)$&
		$\left(
			\begin{array}{cccc}
			    \frac{-1}{5}                      & 0                                 &   0                                  & 0  \\
				0                                  & \frac{-5}{9}                     &   \frac{-8\sqrt{7/5}}{45}            &\frac{4\sqrt{2/105}}{5}  \\
				0                                  & \frac{-8\sqrt{7/5}}{45}           &   \frac{-286}{1575}                  &\frac{-16\sqrt{2/3}}{175}  \\
                0                                  & \frac{4\sqrt{2/105}}{5}           &  \frac{-16\sqrt{2/3}}{175}            &\frac{7}{25}\\
			\end{array}
			\right)$&
		$\left(
			\begin{array}{ccccc}
			\frac{-38}{75}             & \frac{-8}{25\sqrt{21}}           &0                      &0                                 & 0                     \\
             \frac{-8}{25\sqrt{21}}   & \frac{-3}{175}                   &0                      &0                                 & 0                      \\
             0                        & 0                                & \frac{-5}{9}         & \frac{4\sqrt{2}}{45}             & \frac{4\sqrt{2}}{35}   \\
             0                        & 0                                &\frac{4\sqrt{2}}{45}   & \frac{4}{1575}                   & \frac{-4}{175}        \\
             0                        & 0                                &\frac{4\sqrt{2}}{35}   & \frac{-4}{175}                   & \frac{97}{175}       \\
			\end{array}
			\right)$&
		$\left(\begin{array}{cc}
				\frac{-62}{135}                  & \frac{-4}{15\sqrt{11}}   \\
			    \frac{-4}{15\sqrt{11}}           &  \frac{89}{495}          \\
			\end{array}\right) $\\
   $\vec{\epsilon}_{1}\cdot\vec{\epsilon}^{\dagger}_{3}$&$\left(
		\begin{array}{ccc}
			1      &   0    & 0   \\
			0      &   1    & 0  \\
            0      &   0    & 1   \\
		\end{array}
		\right)$&
		$\left(
		\begin{array}{cccc}
			1                     &   0       & 0            & 0    \\
			0                     &   1       & 0            &0    \\
            0                     &   0       & 1            & 0  \\
            0                     &   0       & 0            & 1    \\
		\end{array}
		\right)$&
		$\left(
		\begin{array}{ccccc}
			1     & 0      &  0     &  0  & 0  \\
            0     & 1      &  0     &  0  & 0  \\
            0     & 0      & 1      &  0  & 0  \\
            0      & 0      & 0      &  1  & 0  \\
            0      & 0      & 0      &  0  & 1  \\
		\end{array}
		\right)$&$\left(
			\begin{array}{cc}
				1        & 0  \\
				0  &  1          \\
			\end{array}
			\right)$\\
$iS(\hat{r},\vec{\epsilon}_3^{\dagger}\times\vec{\sigma},\vec{\epsilon}_1)$	
             &$\left(
		\begin{array}{ccc}
			\frac{-1}{35}                         &   \frac{-2\sqrt{2/7}}{5}           & \frac{-12}{35\sqrt{7}}   \\
			\frac{36\sqrt{2/7}}{35}              &   \frac{-76}{2835}                   & \frac{-58\sqrt{2}}{945}      \\
            \frac{-4}{7\sqrt{7}}              &   \frac{34\sqrt{2}}{189}           & \frac{-17}{315}        \\
		\end{array}
		\right)$
            & $\left(
			\begin{array}{cccc}
			    0                       & \frac{3\sqrt{3/5}}{5}             &   \frac{-\sqrt{21}}{25}     & \frac{-8\sqrt{2/7}}{25}  \\
				\frac{-\sqrt{3/5}}{5}     & \frac{8}{63}                     &   \frac{311}{315\sqrt{35}}    &\frac{26\sqrt{2/105}}{35}  \\
				\frac{22\sqrt{3/7}}{25}   & \frac{2\sqrt{5/7}}{63}           &   \frac{-464}{11025}         &\frac{-194\sqrt{2/3}}{1225}  \\
                \frac{-2\sqrt{2/7}}{25} & -2\sqrt{\frac{2}{105}}             &\frac{421\sqrt{2/3}}{1225}   &\frac{46}{1225}              \\
			\end{array}
			\right)$
           & $\left(
			\begin{array}{ccccc}
		  \frac{16}{375}            & \frac{46}{125\sqrt{21}}         &\frac{82}{175\sqrt{3}}  &\frac{-284\sqrt{2/3}}{875}  &\frac{-36\sqrt{6}}{875}           \\
          \frac{121}{125\sqrt{21}}     & \frac{-24}{875}              &\frac{-13}{175\sqrt{7}}   &\frac{-22\sqrt{2/7}}{125} &\frac{-198\sqrt{2/7}}{875}\\
          \frac{27\sqrt{3}}{175}  & \frac{71}{175\sqrt{7}}           &\frac{8}{63}            &\frac{67\sqrt{2}}{1575}            &\frac{-2\sqrt{2}}{525}           \\
          \frac{-68\sqrt{6}}{875}   &\frac{626\sqrt{2/7}}{875} &\frac{-86\sqrt{2}}{1575}   &\frac{404}{7875}     & \frac{-334}{2625}                 \\
           \frac{9\sqrt{6}}{875}    & \frac{-288\sqrt{2/7}}{875} &\frac{-34\sqrt{2}}{525}    &\frac{691}{2625}               & \frac{242}{2625}               \\
			\end{array}
			\right)$
        &$\left(\begin{array}{cc}
				\frac{28}{405}                  & \frac{146}{945\sqrt{11}}   \\
			    \frac{491}{945\sqrt{11}}         & \frac{-62}{495}           \\
			\end{array}\right) $\\
		\hline
       $VB\to{}VB$   &$11/2^{+}$    &$1/2^{+}$                            &$9/2^{-}$ \\
       ${\cal{Z}}$	 &($^{6}F_{J}$) &($^{2}P_{J},^{4}P_{J}, ^{6}F_{J}$)   &($^{6}D_{J}$) \\
       $\vec{\sigma}\cdot{}\vec{L}$ &
		$\left(
		\begin{array}{c}
			15        \\
		\end{array}
		\right)$&
		$\left(
		\begin{array}{ccc}
			-2        &   1  & 0     \\
			1       &   -5    & 0    \\
            0       &   0     & -20  \\
		\end{array}
		\right)$&
		$\left(
		\begin{array}{c}
			10                   \\
		\end{array}
		\right)$\\
     $i\vec{\sigma}\cdot(\vec{\epsilon}_1\times\epsilon_3^{\dagger})$ &$\left(
		\begin{array}{c}
			-1/11                  \\
		\end{array}
		\right)$ &
		$\left(
		\begin{array}{ccc}
			-5/9             &  0                & 0   \\
			0      &   -2/9    & 0  \\
            0     &  0   & 1/3   \\
		\end{array}
		\right)$&
		$\left(
		\begin{array}{c}
			-1/9                   \\
		\end{array}
		\right)$ \\
 $\vec{\epsilon}_{1}\cdot\vec{\epsilon}^{\dagger}_{3}$&
		$\left(
		\begin{array}{c}
			1        \\
		\end{array}
		\right)$&
         $\left(
		\begin{array}{ccc}
			1      &   0    & 0   \\
			0      &   1    & 0  \\
            0      &   0    & 1   \\
		\end{array}
		\right)$&
		$\left(
		\begin{array}{c}
			1        \\
		\end{array}
		\right)$\\
$iS(\hat{r},\vec{\epsilon}_3^{\dagger}\times\vec{\sigma},\vec{\epsilon}_1)$	
             &$\left(
		\begin{array}{c}
			-1/33       \\
		\end{array}
		\right)$
  &$\left(
		\begin{array}{ccc}
			0                             &   \frac{-13}{45\sqrt{5}}  & \frac{-7}{15\sqrt{5}}\\
		\frac{-22}{45\sqrt{5}}            &   \frac{8}{45}            & \frac{-2}{15}        \\
        \frac{17}{15\sqrt{5}}              &   \frac{1}{15}           & \frac{4}{15}        \\
		\end{array}
		\right)$
  &$\left(
		\begin{array}{c}
			-2/63      \\
		\end{array}
		\right)$\\
        \hline
        \hline
	\end{tabular}
\end{table*}

\begin{table*}[t!]
	\centering
	\caption{The matrix elements of two-body interaction operators for $VB$ systems.}\label{tableqw-1}
     \setlength{\tabcolsep}{1.5pt}
	\begin{tabular}{cccccccc}
		\hline\hline
		$VB\to{}VB$    &$1/2^{-}$                    & $3/2^{-}$                               &$3/2^{+}$                  &$7/2^{-}$         \\
${\cal{Z}}$		&($^{2}S_{J}$,$^{4}D_{J}$, $^{6}D_{J}$)& ($^{4}S_{J},{}^{2}D_{J},{}^{4}D_{J},{}^{6}D_{J}$) &($^2P_{J},{}^{4}P_{J}, {}^{6}P_{J},{}^{4}F_{J}, {}^{6}F_{J}$)
&($^4D_{J}, ^{6}D_J$)   \\
$iS(\hat{r},\vec{\sigma},\vec{\epsilon}_1\times\epsilon_3^{\dagger})$	
             &$\left(
		\begin{array}{ccc}
			0                             &   \frac{7}{9\sqrt{5}}    & \frac{-2}{3\sqrt{5}}\\
		\frac{7}{9\sqrt{5}}               &   \frac{-16}{45}         & \frac{1}{15}        \\
        \frac{-2}{3\sqrt{5}}              &   \frac{1}{15}           & \frac{-8}{15}        \\
		\end{array}
		\right)$
            & $\left(
			\begin{array}{cccc}
			    0                                   & \frac{-41\sqrt{\frac{2}{5}}}{45}   &   \frac{4}{9}                &\frac{\sqrt{14}}{15}   \\
				\frac{-41\sqrt{\frac{2}{5}}}{45}            & \frac{-2}{9}               &   \frac{\sqrt{\frac{2}{5}}}{45}      &\frac{4}{5\sqrt{35}}  \\
				\frac{4}{9}                         & \frac{\sqrt{\frac{2}{5}}}{45}      &   \frac{-32}{225}            &\frac{-\sqrt{14}}{75}  \\
                \frac{\sqrt{14}}{15}                & \frac{4}{5\sqrt{35}}       &   \frac{-\sqrt{14}}{75}      &\frac{-242}{525}          \\
			\end{array}
			\right)$
           & $\left(
			\begin{array}{ccccc}
			\frac{-2}{9}                    & \frac{-13\sqrt{2}}{75}       &\frac{44}{75\sqrt{3}}       &\frac{19\sqrt{2}}{75}    & \frac{-8\sqrt{2}}{75}    \\
			\frac{-13\sqrt{2}}{75}          & \frac{176}{1125}             &\frac{107\sqrt{\frac{2}{5}}}{375}   &\frac{164}{375}          & \frac{32}{375}           \\
            \frac{44}{75\sqrt{3}}           & \frac{107\sqrt{\frac{2}{5}}}{375}    &\frac{-34}{125}             &\frac{-9\sqrt{6}}{125}   & \frac{8\sqrt{6}}{125}    \\
            \frac{19\sqrt{2}}{75}           & \frac{164}{375}              &\frac{-9\sqrt{6}}{125}      &\frac{-112}{375}         & \frac{-56}{375}          \\
            \frac{-8\sqrt{2}}{75}           & \frac{32}{375}               &\frac{8\sqrt{6}}{125}       &\frac{-56}{375}          & \frac{-178}{375}         \\
			\end{array}
			\right)$
            &$\left(\begin{array}{cc}
				\frac{-88}{735}                    & \frac{-218\sqrt{\frac{2}{3}}}{735}   \\
			    \frac{-218\sqrt{\frac{2}{3}}}{735}         & \frac{454}{2205}             \\
			\end{array}\right) $\\
$iS(\hat{r},\vec{\epsilon}_1\times\vec{\sigma},\vec{\epsilon}_3^{\dagger})$	
        &$\left(
		\begin{array}{ccc}
			0                             &   \frac{22}{45\sqrt{5}}  & \frac{-17}{15\sqrt{5}}\\
		\frac{13}{45\sqrt{5}}             &   \frac{-8}{45}          & \frac{-1}{15}        \\
        \frac{7}{15\sqrt{5}}              &   \frac{2}{15}           & \frac{-4}{15}        \\
		\end{array}
		\right)$
       & $\left(
			\begin{array}{cccc}
			    0                                   & \frac{-22\sqrt{\frac{2}{5}}}{45}   &   \frac{56}{225}                    &\frac{16\sqrt{\frac{2}{7}}}{75}     \\
			\frac{-19\sqrt{\frac{2}{5}}}{45}        & \frac{-1}{9}                       &   \frac{2\sqrt{10}}{63}             &\frac{8}{7\sqrt{35}}                \\
				\frac{4}{225}                       & \frac{-43\sqrt{\frac{2}{5}}}{315}  &   \frac{-16}{225}                   &\frac{-227\sqrt{\frac{2}{7}}}{525}  \\
                \frac{19\sqrt{\frac{2}{7}}}{75}     & \frac{-12}{35\sqrt{35}}            &   \frac{178\sqrt{\frac{2}{7}}}{525} &\frac{-121}{525}          \\
			\end{array}
			\right)$
      & $\left(
			\begin{array}{ccccc}
			\frac{-1}{9}                    & \frac{-2\sqrt{2}}{75}       &\frac{76}{75\sqrt{3}}             &\frac{97\sqrt{2}}{525}    & \frac{-79\sqrt{2}}{525}    \\
			\frac{-11\sqrt{2}}{75}          & \frac{88}{1125}             &\frac{-59\sqrt{\frac{2}{5}}}{375} &\frac{884}{2625}          & \frac{-8}{2625}           \\
            \frac{-32}{75\sqrt{3}}          & \frac{166\sqrt{\frac{2}{3}}}{375}    &\frac{-17}{125}             &\frac{-29\sqrt{6}}{875}   & \frac{23\sqrt{6}}{875}    \\
            \frac{12\sqrt{2}}{175}           & \frac{88}{875}              &\frac{-34\sqrt{6}}{875}      &\frac{-56}{375}         & \frac{-496}{2625}          \\
            \frac{23\sqrt{2}}{525}           & \frac{232}{2625}               &\frac{33\sqrt{6}}{875}       &\frac{104}{2625}          & \frac{-89}{375}         \\
			\end{array}
			\right)$
      &$\left(\begin{array}{cc}
				\frac{-44}{735}                           & \frac{-166\sqrt{\frac{2}{3}}}{735}   \\
			    \frac{-52\sqrt{\frac{2}{3}}}{735}         & \frac{227}{2205}             \\
			\end{array}\right) $\\
		\hline
   $VB\to{}VB$    &$7/2^{+}$                    & $5/2^{-}$                               &$5/2^{+}$                   &$9/2^{+}$                   \\
${\cal{Z}}$		&($^{6}P_{J},^{4}F_{J},^{6}F_{J}$)& ($^{6}S_{J},^{2}D_{J},^{4}D_{J},^{6}D_{J}$) &($^{4}P_J, ^{6}P_{J}, ^{2}F_{J}, ^{4}F_{J}, ^{6}F_J$) &($^{4}F_{J},^{6}F_{J}$) \\
$iS(\hat{r},\vec{\sigma},\vec{\epsilon}_1\times\epsilon_3^{\dagger})$	
             &$\left(
		\begin{array}{ccc}
			\frac{2}{35}                          &  \frac{-22\sqrt{\frac{2}{7}}}{35}    & \frac{32}{35\sqrt{7}}        \\
			\frac{-22\sqrt{\frac{2}{7}}}{35}      &   \frac{152}{2835}                   & \frac{-16\sqrt{2}}{135}      \\
            \frac{32}{35\sqrt{7}}                 &  \frac{-16\sqrt{2}}{135}             & \frac{34}{315}              \\
		\end{array}
		\right)$
            & $\left(
			\begin{array}{cccc}
			    0                          & \frac{-2\sqrt{\frac{3}{5}}}{5}   &\frac{-3\sqrt{\frac{3}{7}}}{5}      & \frac{2\sqrt{\frac{2}{7}}}{5}      \\
			\frac{-2\sqrt{\frac{3}{5}}}{5} & \frac{-16}{63}                   &\frac{-361}{315\sqrt{35}}           & \frac{44\sqrt{\frac{2}{105}}}{35}          \\
			\frac{-3\sqrt{\frac{3}{7}}}{5} & \frac{-361}{315\sqrt{35}}        &\frac{928}{11025}                   &\frac{-227\sqrt{\frac{2}{3}}}{1225} \\
            \frac{2\sqrt{\frac{2}{7}}}{5}  & \frac{44\sqrt{\frac{2}{105}}}{35}&\frac{-227\sqrt{\frac{2}{3}}}{1225} &\frac{-92}{1225}                    \\
			\end{array}
			\right)$
           & $\left(
			\begin{array}{ccccc}
		  \frac{-32}{375}            & \frac{-167}{125\sqrt{21}}         &\frac{-163}{175\sqrt{3}}  &\frac{488\sqrt{\frac{2}{3}}}{875}  &\frac{27\sqrt{6}}{875}           \\
          \frac{-167}{125\sqrt{21}}  & \frac{48}{875}                    &\frac{-58}{175\sqrt{7}}   &\frac{-472\sqrt{\frac{2}{7}}}{875} &\frac{486\sqrt{\frac{2}{7}}}{875}\\
          \frac{-163}{175\sqrt{3}}   & \frac{-58}{175\sqrt{7}}           &\frac{-16}{63}            &\frac{19\sqrt{2}}{1575}            &\frac{12\sqrt{2}}{175}           \\
   \frac{488\sqrt{\frac{2}{3}}}{875} &\frac{-472\sqrt{\frac{2}{7}}}{875} &\frac{19\sqrt{2}}{1575}   &\frac{-808}{7875}                  & \frac{-17}{125}                 \\
           \frac{27\sqrt{6}}{875}    & \frac{486\sqrt{\frac{2}{7}}}{875} &\frac{12\sqrt{2}}{175}    &\frac{-17}{125}                    & \frac{-484}{2625}               \\
			\end{array}
			\right)$
        &$\left(\begin{array}{cc}
				\frac{-56}{405}                  & \frac{-91}{135\sqrt{11}}   \\
			    \frac{-91}{135\sqrt{11}}         & \frac{124}{495}           \\
			\end{array}\right) $\\	
$iS(\hat{r},\vec{\epsilon}_1\times\vec{\sigma},\vec{\epsilon}_3^{\dagger})$	
        &$\left(
		\begin{array}{ccc}
		\frac{1}{35}                              &   \frac{-36\sqrt{\frac{2}{7}}}{35}     & \frac{4}{7\sqrt{5}}            \\
		\frac{2\sqrt{\frac{2}{7}}}{5}             &   \frac{76}{2835}                      & \frac{-34\sqrt{2}}{189}        \\
        \frac{12}{35\sqrt{7}}                     &   \frac{58\sqrt{2}}{945}               & \frac{17}{315}        \\
		\end{array}
		\right)$
& $\left(
			\begin{array}{cccc}
			    0                          & \frac{\sqrt{\frac{3}{5}}}{5}   &\frac{-22\sqrt{\frac{3}{7}}}{25}      & \frac{2\sqrt{\frac{2}{7}}}{25}      \\
			\frac{-3\sqrt{\frac{3}{5}}}{5} & \frac{-8}{63}                   &\frac{-2\sqrt{\frac{5}{7}}}{63}      & 2\sqrt{\frac{2}{105}}          \\
			\frac{\sqrt{21}}{25}           & \frac{-311}{315\sqrt{35}}        &\frac{464}{11025}                   &\frac{-421\sqrt{\frac{2}{3}}}{1225} \\
            \frac{8\sqrt{\frac{2}{7}}}{25}  & \frac{-26\sqrt{\frac{2}{105}}}{35}&\frac{194\sqrt{\frac{2}{3}}}{1225} &\frac{-46}{1225}                    \\
			\end{array}
			\right)$
& $\left(
			\begin{array}{ccccc}
		  \frac{-16}{375}            & \frac{-121}{125\sqrt{21}}         &\frac{-27\sqrt{3}}{175}  &\frac{68\sqrt{6}}{875}  &\frac{-9\sqrt{6}}{875}           \\
          \frac{-46}{125\sqrt{21}}     & \frac{24}{875}                    &\frac{-71}{175\sqrt{7}}   &\frac{-626\sqrt{\frac{2}{7}}}{875} &\frac{288\sqrt{\frac{2}{7}}}{875}\\
          \frac{-82}{175\sqrt{3}}  & \frac{13}{175\sqrt{7}}           &\frac{-8}{63}            &\frac{86\sqrt{2}}{1575}            &\frac{34\sqrt{2}}{525}           \\
          \frac{284\sqrt{\frac{2}{3}}}{875}   &\frac{22\sqrt{\frac{2}{7}}}{125} &\frac{-67\sqrt{2}}{1575}   &\frac{-404}{7875}     & \frac{-691}{2625}                 \\
           \frac{36\sqrt{6}}{875}    & \frac{198\sqrt{\frac{2}{7}}}{875} &\frac{2\sqrt{2}}{525}    &\frac{334}{2625}               & \frac{-242}{2625}               \\
			\end{array}
			\right)$
&$\left(\begin{array}{cc}
				\frac{-28}{405}                  & \frac{-491}{945\sqrt{11}}   \\
			    \frac{-146}{945\sqrt{11}}         & \frac{62}{495}           \\
			\end{array}\right) $\\
		\hline
       $VB\to{}VB$   &$11/2^{+}$    &$1/2^{+}$                            &$9/2^{-}$ \\
       ${\cal{Z}}$	 &($^{6}F_{J}$) &($^{2}P_{J},^{4}P_{J}, ^{6}F_{J}$)   &($^{6}D_{J}$) \\
$iS(\hat{r},\vec{\sigma},\vec{\epsilon}_1\times\epsilon_3^{\dagger})$		
             &$\left(
		\begin{array}{c}
			\frac{2}{33}       \\
		\end{array}
		\right)$
    &$\left(
		\begin{array}{ccc}
			0                     &\frac{7}{9\sqrt{5}}           &\frac{-2}{3\sqrt{5}} \\
			\frac{7}{9\sqrt{5}}   &\frac{-16}{45}                &\frac{1}{15}         \\
            \frac{-2}{3\sqrt{5}}  &\frac{1}{15}                  &\frac{-8}{15}        \\
		\end{array}
		\right)$
    & $\left(
		\begin{array}{c}
			\frac{4}{63}       \\
		\end{array}
		\right)$\\
 $iS(\hat{r},\vec{\epsilon}_1\times\vec{\sigma},\vec{\epsilon}_3^{\dagger})$
 &$\left(
		\begin{array}{c}
			\frac{1}{33}       \\
		\end{array}
		\right)$
 &$\left(
		\begin{array}{ccc}
			0                             &   \frac{22}{45\sqrt{5}}  & \frac{-17}{15\sqrt{5}}\\
		\frac{13}{45\sqrt{5}}             &   \frac{-8}{45}          & \frac{-1}{15}        \\
        \frac{7}{15\sqrt{5}}              &   \frac{2}{15}           & \frac{-4}{15}        \\
		\end{array}
		\right)$
& $\left(
		\begin{array}{c}
			\frac{2}{63}       \\
		\end{array}
		\right)$\\
        \hline
        \hline
	\end{tabular}
\end{table*}

Since the isospin of $N(2250)$ and $\Delta(2200)$ baryons are $1/2$ and $3/2$, respectively,  the flavor wavefunction $|I, I_3\rangle$ for the $K^{*}\Sigma^{*}$
systems is simple and reads as
\begin{align}
|N^{*+}\rangle = \sqrt{\frac{1}{3}}|K^{*+} \Sigma^{*0}\rangle-\sqrt{\frac{2}{3}}|K^{*0} \Sigma^{*+}\rangle, \nonumber\\
|\Delta^{*+}\rangle = \sqrt{\frac{2}{3}}|K^{*+} \Sigma^{*0}\rangle+\sqrt{\frac{1}{3}}|K^{*0} \Sigma^{*+}\rangle\label{eq28},
\end{align}
where $I$ and $I_3$ are the isospin and its third component. With the isospin configuration of the system, the total interaction potential can be easily obtained;
it is omitted here for brevity.

\section{RESULTS AND DISCUSSIONS}\label{Sec: results}
Using the obtained effective interaction potentials, we investigate the possible $K^{*}\Sigma^{*}$ bound states by solving the nonrelativistic Schr\"{o}dinger equation with
the Gaussian expansion method (GEM). The numerical results are summarized in Table~\ref{c4}.  The calculations are performed for both isospin configurations, $I=1/2$ and $I=3/2$, corresponding to the nucleon $N$ and $\Delta$ resonance channels, respectively, with the spin-parity quantum numbers of the considered states listed in Table~\ref{tab:spins}.
Since the coupling constant $H_{A}$ can take the values $2.27$ and $1.74$, we also present the corresponding results for these two choices.

In the present model, $\alpha$ is a free parameter associated with the form factor and cannot be determined from first principles. Phenomenological studies typically constrain it to the range $0.5$--$9.7$~\cite{Dong:2017rmg,Liu:2006df,Xu:2015qqa,Chen:2012nva,Jian:2022rln}. For example, $\alpha=0.5$--$5.0$ provides a reasonable description of the decays $X(3872)\to J/\psi\rho$ and $Y(3940)\to J/\psi\omega$~\cite{Liu:2006df}, while fitting the $e^+e^- \to \bar{p}p\pi^0$ data at $\sqrt{s}=3.773~\mathrm{GeV}$ yields $\alpha=6.2\pm3.5$~\cite{Xu:2015qqa}. Moreover, the branching fractions of $\psi(4040)\to J/\psi\eta$ and $\psi(4160)\to J/\psi\eta$ can be reproduced with $\alpha=0.53$--$1.20$~\cite{Chen:2012nva}. Therefore, we vary $\alpha$ from $0.5$ to $9.7$ to search for possible bound-state solutions.

As shown in Table~\ref{c4}, for the isospin $I=1/2$ and $J^{P}=1/2^{-}$ channel, which corresponds to the $S$-wave $K^{*}\Sigma^{*}$  molecular configuration, no stable bound-state
solution is found. Specifically, for both values of the coupling constant, $H_{A}=2.27$ and $H_{A}=1.74$, and for all physically reasonable values of the parameter $\alpha$ in the
range $0.5$--$9.7$~\cite{Dong:2017rmg,Liu:2006df,Xu:2015qqa,Chen:2012nva,Jian:2022rln}, no negative and stable energy eigenvalue is obtained.  This indicates that such a bound state does not exist in this
channel.  In contrast,  a bound state is found in the $I=3/2$, $J^{P}=1/2^{-}$ channel, also corresponding to the $S$-wave $K^{*}\Sigma^{*}$ molecular configuration.  This state emerges
at $\alpha = 2.0$ with a binding energy of $E = -0.705$~MeV.  As $\alpha$ further increases (with a step size of 0.5 in our calculation), the binding energy becomes larger, indicating
that the $K^{*}\Sigma^{*}$ molecular state becomes more deeply bound.

No bound state is found in the $I=1/2,,J^{P}=1/2^-$ channel, whereas a bound state emerges in the $I=3/2,,J^{P}=1/2^-$ channel. This difference originates from the strong isospin
dependence of the $K^{*}\Sigma^{*}$ interaction. Using Eq.~(\ref{eq28}), the difference between the interaction potentials in the two isospin channels can be written as
$V(\Delta^{*})-V(N^{*})
=\frac{1}{3}V_{K^{*+}\Sigma^{*0}\to K^{*+}\Sigma^{*0}}
-\frac{1}{3}V_{K^{*0}\Sigma^{*+}\to K^{*0}\Sigma^{*+}}
+\frac{2\sqrt{2}}{3}V_{K^{*+}\Sigma^{*0}\to K^{*0}\Sigma^{*+}}$.  As a result, the attraction in the $I=1/2$ channel is
insufficient to generate a bound state, whereas it is enhanced in the $I=3/2$ channel and becomes strong enough to support binding.
\begin{figure}[http]
\begin{center}
\includegraphics[bb=-120 10 750 395, clip, scale=0.30]{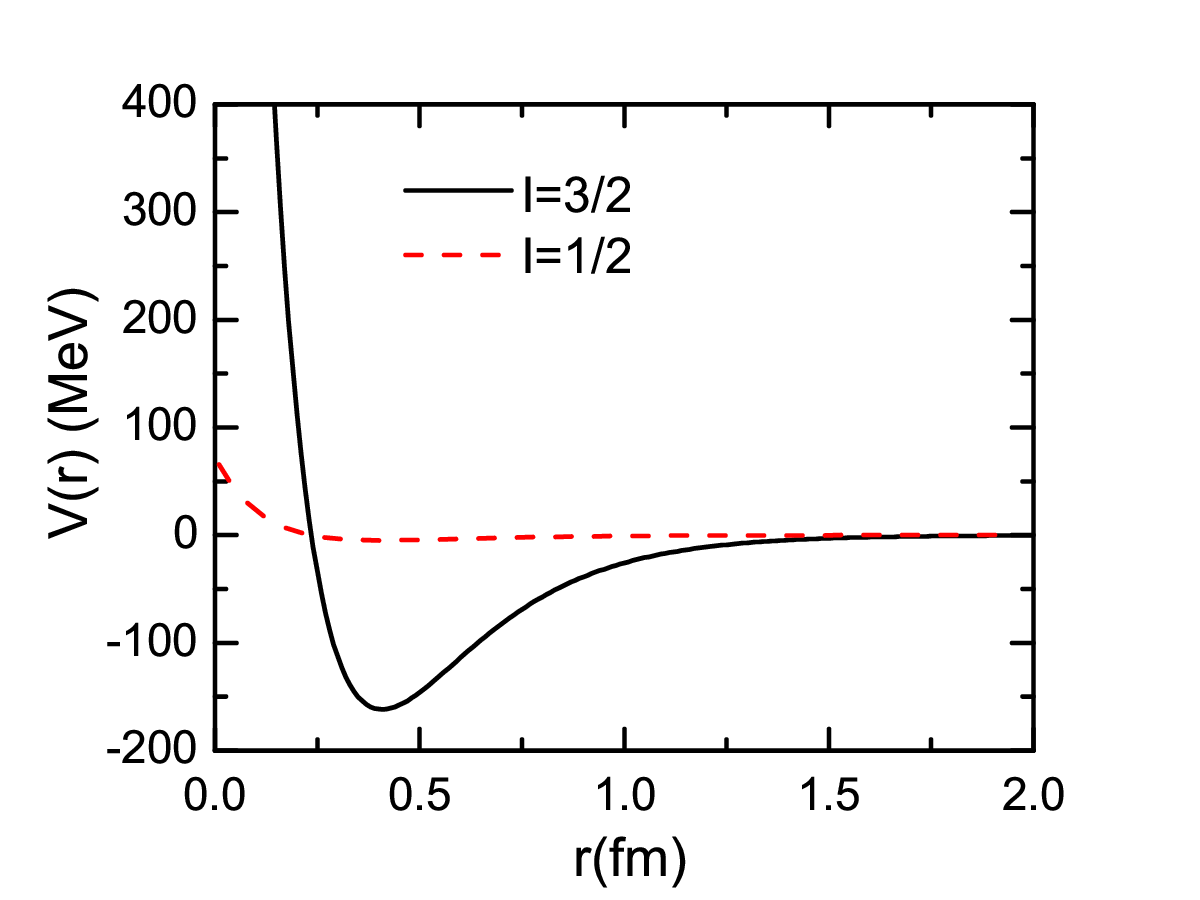}
\caption{Interaction potentials in different isospin sectors for the $J^P=1/2^{-}$ channel.}\label{fig:shi}
\end{center}
\end{figure}

Indeed, the interaction potentials for the $^{2}S_{1/2}\to {}^{2}S_{1/2}$ partial wave are shown in Fig.~\ref{fig:shi} as supporting evidence.  It is found that the $I=3/2$ channel exhibits a
significantly stronger attraction, reaching up to $-161.79~\text{MeV}$ at $\alpha=2.0$, where a bound state just begins to emerge. This value is approximately 34 times larger than that of the
$I=1/2$ channel, which is only $-4.76~\text{MeV}$.   This isospin-dependent interaction plays an important role in facilitating the formation of bound states, a feature that can be also observed
in other channels. A direct indication is
that the $\alpha$ values required to generate bound states in the $I=1/2$ channel are systematically larger than those in the $I=3/2$ channel (see Tab.~\ref{c4}).  The results also indicate that
the two $H_{A}$ values considered in the present work have no significant impact on the final results, which remain nearly identical across all $J^{P}$ channels, with no appreciable changes in
the bound-state properties upon their inclusion.
\begin{table*}[htbp!]
\centering
\setlength{\tabcolsep}{8pt}
\caption{Possible bound states for the $K^{*}\Sigma^{*}$ interaction with $\Lambda_i = m_i + 220 \alpha~\mathrm{MeV}$ and different spin-parity assignments.
$E$ denote the eigenvalues (in units of MeV). The symbol $\times$ indicates that no bound-state solution is found.}\label{c4}
\begin{tabular}{ccccccccccccccccccccc}
\hline\hline
          &               &\multicolumn{2}{c}{$H_A=2.27$}  &~~&\multicolumn{2}{c}{$H_A=1.74$} &               &\multicolumn{2}{c}{$H_A=2.27$} &~~&\multicolumn{2}{c}{$H_A=1.74$}\\
                          \cline{3-4}                       \cline{6-7}                                       \cline{9-10}                    \cline{12-13}
Isospin   &State          &$\alpha$       &$E$             &~~&$\alpha$      &$E$             &State          &$\alpha$       &$E$            &~~&$\alpha$      &$E$         \\\hline
$I=1/2$   &$J^P=1/2^{-}$  &$\times$       &$\times$        &~~&$\times$      &$\times$        &$J^P=1/2^+$    &$\times$       &$\times$       &~~& $\times$     &$\times$    \\
          &               &$\times$       &$\times$        &~~&$\times$      &$\times$        &               &$\times$       &$\times$       &~~& $\times$     &$\times$    \\
          &               &$\times$       &$\times$        &~~&$\times$      &$\times$        &               &$\times$       &$\times$       &~~& $\times$     &$\times$    \\
          &$J^P=3/2^{-}$  &$4.4$          &$-1.374$        &~~&$4.4$         &$-1.374$        &$J^P=3/2^+$    &$6.3$          &$-1.003$       &~~& $6.3$        &$-1.002$    \\
          &               &$4.9$          &$-8.401$        &~~&$4.9$         &$-8.401$        &               &$6.8$          &$-29.844$      &~~& $6.8$        &$-29.844$    \\
          &               &$5.4$          &$-20.946$       &~~&$5.4$         &$-20.945$       &               &$7.3$          &$-65.938$      &~~& $7.3$        &$-65.937$    \\
          &$J^P=5/2^{-}$  &$3.9$          &$-0.119$        &~~&$3.9$         &$-0.119$        &$J^P=5/2^+$    &$4.0$          &$-2.471$       &~~& $4.0$        &$-2.471$    \\
          &               &$4.4$          &$-6.334$        &~~&$4.4$         &$-6.334$        &               &$4.2$          &$-37.680$      &~~& $4.2$        &$-37.680$    \\
          &               &$4.9$          &$-20.827$       &~~&$4.9$         &$-20.826$       &               &$4.4$          &$-65.938$      &~~& $4.4$        &$-65.938$    \\
          &$J^P=7/2^{+}$  &$2.95$         &$-0.882$        &~~&$3.9$         &$-0.882$        &$J^P=7/2^-$    &$4.76$         &$-4.461$       &~~& $4.76$       &$-4.460$    \\
          &               &$3.05$         &$-32.953$       &~~&$3.05$        &$-32.953$       &               &$4.86$         &$-65.981$      &~~& $4.86$       &$-65.980$    \\
          &               &$3.15$         &$-79.136$       &~~&$3.15$        &$-79.136$       &               &$4.96$         &$-143.682$     &~~& $4.96$       &$-143.682$    \\
 -         &$J^P=9/2^{+}$  &$5.51$         &$-0.796$        &~~&$5.51$        &$-0.795$       &$J^P=9/2^-$    &$3.39$         &$-8.344$       &~~& $3.39$       &$-8.344$    \\
          &               &$5.52$         &$-12.042$       &~~&$5.52$        &$-12.041$       &               &$3.40$         &$-17.408$      &~~& $3.40$       &$-17.408$    \\
          &               &$5.53$         &$-23.562$       &~~&$5.53$        &$-23.561$       &               &$3.41$         &$-26.720$      &~~& $3.41$       &$-21.720$    \\
          &$J^P=11/2^{+}$ &$\times$       &$\times$        &~~&$\times$       &$\times$            \\
          &               &$\times$       &$\times$        &~~&$\times$       &$\times$           \\
          &               &$\times$       &$\times$        &~~&$\times$       &$\times$           \\
\hline
\hline
$I=3/2$   &$J^P=1/2^{-}$  &2.0            &$-0.705$        &~~& 2.0          &$-0.705$        &$J^P=1/2^+$    &$\times$       &$\times$       &~~& $\times$     &$\times$    \\
          &               &2.5            &$-16.577$       &~~& 2.5          &$-16.576$       &               &$\times$       &$\times$       &~~& $\times$     &$\times$    \\
          &               &3.0            &$-52.089$       &~~& 3.0          &$-52.088$       &               &$\times$       &$\times$       &~~& $\times$     &$\times$     \\
          &$J^p=3/2^{-}$  &1.9            &$-0.821$        &~~&1.9           &$-0.821$        &$J^P=3/2^+$    &$2.6$          &$-1.903$       &~~& $2.6$        &$-1.903$    \\
          &               &2.4            &$-18.988$       &~~&2.4           &$-18.988$       &               &$3.1$          &$-70.925$      &~~& $3.1$        &$-70.925$    \\
          &               &2.9            &$-58.968$       &~~&2.9           &$-58.968$       &               &$3.6$          &$-217.168$     &~~& $3.6$        &$-217.66$   \\
          &$J^p=5/2^{-}$  &1.8            &$-1.125$        &~~&1.9           &$-0.821$        &$J^P=5/2^+$    &$1.75$         &$-0.990$       &~~& $1.75$       &$-0.990$    \\
          &               &2.3            &$-26.779$       &~~&2.4           &$-18.988$       &               &$1.85$         &$-31.237$      &~~& $1.85$       &$-31.237$    \\
          &               &2.8            &$-83.566$       &~~&2.9           &$-58.968$       &               &$1.95$         &$-84.207$      &~~& $1.95$       &$-84.206$   \\
          &$J^P=7/2^{+}$  &$1.34$         &$-0.740$        &~~&$1.34$        &$-0.740$        &$J^P=7/2^-$    &$2.04$         &$-5.062$       &~~& $2.04$       &$-5.061$    \\
          &               &$1.39$         &$-29.323$       &~~&$1.39$        &$-29.323$       &               &$2.05$         &$-14.090$      &~~& $2.05$       &$-14.090$    \\
          &               &$1.44$         &$-71.788$       &~~&$1.44$        &$-71.788$       &               &$2.06$         &$-23.776$      &~~& $2.06$       &$-23.776$    \\
          &$J^P=9/2^{+}$  &$2.74$         &$\times$        &~~&$2.74$        &$\times$        &$J^P=9/2^-$    &$2.16$         &$-0.398$       &~~& $2.16$       &$-0.398$    \\
          &               &$2.75$         &$-2.887$        &~~&$2.75$        &$-2.887$        &               &$2.17$         &$-12.665$      &~~& $2.17$       &$-12.66$    \\
          &               &$2.76$         &$-18.876$       &~~&$2.76$        &$-18.876$       &               &$2.18$         &$-25.787$      &~~& $2.18$       &$-25.787$    \\
          &$J^P=11/2^{+}$ &$8.29$         &$\times$        &~~&$8.29$        &$\times$            \\
          &               &$8.30$         &$-4.429$        &~~&$8.30$        &$-4.429$           \\
          &               &$8.31$         &$-16.888$       &~~&$8.31$        &$-16.888$           \\
\hline \hline
\end{tabular}
\end{table*}

For the $J^{P}=3/2^{-}$ state, we consider the mixing of $\left|{}^{4}S_{3/2} \sim {}^{2}D_{3/2} \sim {}^{4}D_{3/2} \sim {}^{6}D_{3/2}\right\rangle$. The corresponding numerical results
are summarized in Table~\ref{c4}. We find that a $K^{*}\Sigma^{*}$ bound state emerges at $\alpha = 4.4$, with a binding energy of $1.374\,\mathrm{MeV}$ for the $N$ state, while for the $\Delta$
state a bound state appears at $\alpha = 1.9$, with a binding energy of $0.821\,\mathrm{MeV}$. As the $\alpha$ increases, the binding energy becomes progressively deeper.  These results
indicate that the effective potential provides enough attraction to overcome the centrifugal barrier, leading to the formation of a bound state with a negative energy eigenvalue.

A detailed
analysis shows that the $S$--$D$ wave mixing plays a crucial role in the formation of the bound state.  Taking the case with isospin $I = 3/2$ and $\alpha = 2.4$, where a binding energy of
$E = 18.988\,\mathrm{MeV}$ is obtained, as an example: when only the $S$-wave contribution is considered, a bound state with $E = 4.509\,\mathrm{MeV}$ appears at $\alpha = 2.4$; however, for
the pure $D$-wave case, no bound state can be obtained at the same value of $\alpha$, and a bound state only emerges when $\alpha$ is increased to $\alpha = 7.852$. Furthermore, when the tensor
force is switched off, no bound-state solution can be obtained in the pure $D$-wave channel. These results indicate that although the $S$-wave component provides the dominant contribution, the
$S$--$D$ coupling significantly enhances the effective attraction, and that the tensor interaction plays a key role in stabilizing the bound state.

\begin{figure}[http]
\begin{center}
\includegraphics[bb=-100 8 750 330, clip, scale=0.36]{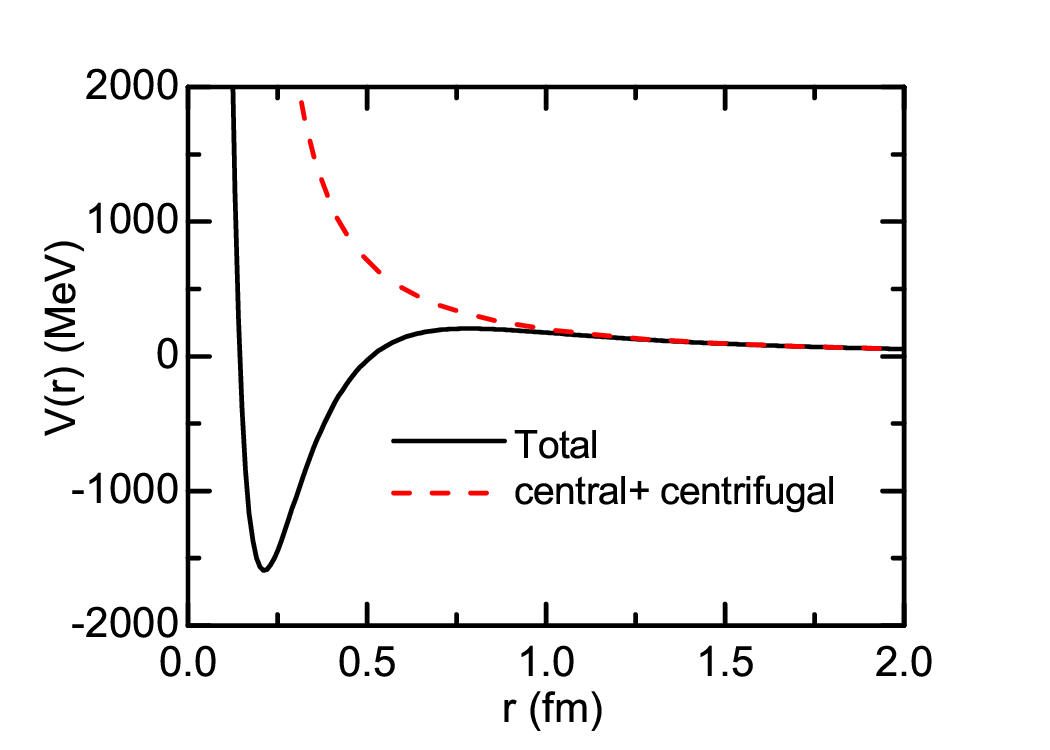}
\caption{Interaction potentials in the $I=1/2$, $J^P=9/2^{-}$ channel as functions of the radial distance $r$ for $\alpha=5.25$. The black solid curve shows the total interaction potential,
whereas the red dashed curve includes only the central potential and the centrifugal potential.}\label{hyug}
\end{center}
\end{figure}

The formation mechanism of the molecular state with $J^{P}=5/2^{-}$ follows the same pattern as that of the $J^{P}=3/2^{-}$ case. This feature extends to states involving different partial-wave
couplings, including $J^{P}=3/2^{+}$, $J^{P}=5/2^{+}$, and $J^{P}=7/2^{+}$, in which the binding is dominated by lower partial waves, with the $P$--$F$ coupling providing an additional enhancement.
By contrast, for the higher partial-wave configurations with $J^{P}=7/2^{-}$ and $J^{P}=9/2^{\pm}$, where partial-wave mixing is negligible, the formation of the bound states is primarily driven by
the attractive contribution from the non-central interaction. As an illustrative example, Fig.~\ref{hyug} presents the interaction potentials in the $I=1/2$, $J^P=9/2^{-}$ channel. It is evident that the total interaction potential (black solid curve) is sufficiently attractive to support a bound state, whereas the interaction potential including only the central potential and the centrifugal potential (red dashed curve) remains repulsive throughout the entire radial range. This comparison indicates that the non-central interaction provides the dominant attractive contribution required to overcome the centrifugal repulsion and generate the bound state in this channel.

The states $N(2250)$ and $\Delta(2200)$ correspond to the $I=\tfrac{1}{2}$ channel with $J^{P}=9/2^{-}$, where only the ${}^{6}D$ contribution is included, and the $I=\tfrac{3}{2}$ channel with $J^{P}=7/2^{-}$,
where both the ${}^{4}D$ and ${}^{6}D$ partial waves are considered. Within the present framework, both states can be understood as molecular states. From Table~\ref{c4}, it can be seen that in the $I=\tfrac{3}{2}$, $J^{P}=7/2^{-}$ channel, a bound state with binding energy $E=5.062~\mathrm{MeV}$ is obtained at $\alpha=2.04$. To reproduce the $\Delta(2200)$ state, the parameter $\alpha$ needs to be slightly increased to $2.10$.
Similarly, a much larger value, $\alpha=3.51$ (not listed in Table~\ref{c4}), is required to interpret the $N(2220)$ as a $K^{*}\Sigma^{*}$ molecular state.

\begin{figure}[http]
\begin{center}
\includegraphics[bb=-20 8 750 280, clip, scale=0.36]{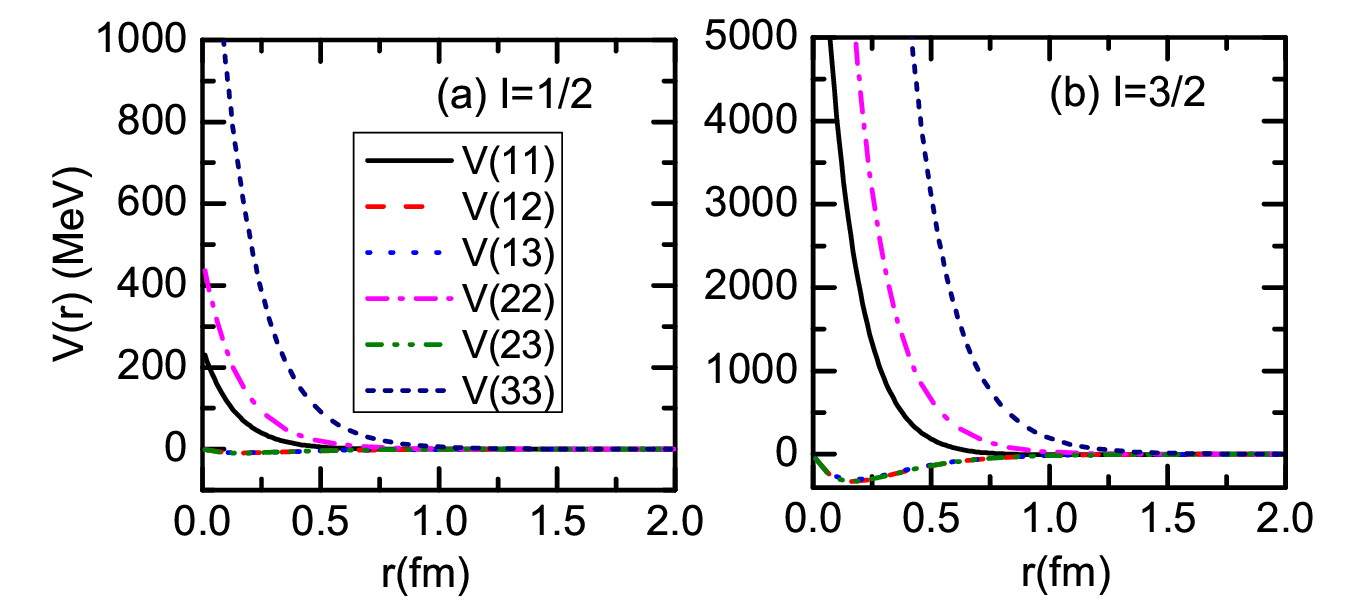}
\caption{Interaction potentials in different isospin sectors for the $J^P=1/2^{+}$ channel. Here, $V(ij)$ ($i,j = 1(^{2}P), 2(^{4}P), 3(^{6}F)$) denotes the interaction potentials between different
partial waves.}\label{fig:shi-12-plus}
\end{center}
\end{figure}
For the remaining two higher partial-wave channels, corresponding to $J^{P}=1/2^{+}$ and $J^{P}=11/2^{+}$, the former involves the coupled components $|{}^{2}P_{1/2}\sim{}^{4}P_{1/2}\sim{}^{6}F_{1/2}\rangle$,
whereas the latter receives contributions purely from the $|{}^{6}F_{11/2}\rangle$ configuration.  We find that for the $J^P = 1/2^{+}$ channel,  no bound $K^{*}\Sigma^{*}$ molecular states are obtained
for either $I=1/2$ or $I=3/2$.  This can be understood from two main aspects. First, the diagonal potentials are strongly repulsive, largely exceeding the off-diagonal attractive interactions,
as illustrated in Fig.~\ref{fig:shi-12-plus}.  Note that only the upper-right part of the potential matrix is presented, as the lower-left part shows similar behavior and has the same order of
magnitude.  Second, a strong centrifugal barrier arises from the nonzero orbital angular momentum, which further suppresses bound-state formation.

Fig.~\ref{fig:shi-12-plus} also tell us that the three diagonal potentials exhibit a pronounced repulsive behavior in the short-range region $r < 0.75\,\mathrm{fm}$,
with magnitudes reaching up to several thousand MeV. This behavior is commonly attributed to the contributions of vector-meson exchange (such as $\rho$ and $\omega$) in the short-distance
limit~\cite{Machleidt:1987hj}, and in the framework of effective field theory is treated as contact terms, whose strength is fitted or encoded in low-energy constants (LECs)~\cite{Weinberg:1990rz}.
This result is consistent with the contributions of the $\rho$ and $\omega$ mesons included in our system, as well as the $\pi$ meson, which provides the long-range interaction.

For the $J^{P}=11/2$ case, a bound state is absent in the $I=1/2$ channel but appears in the $I=3/2$ channel. This contrast originates from the interference pattern of the interaction
potentials in isospin space, as discussed previously.  A comparison between the $J^{P}=1/2^{+}$ and $J^{P}=11/2^{+}$ cases shows that the absence of a bound state in the $I=\tfrac{1}{2}$ channel
for $J^{P}=11/2^{+}$ is due to the meson-exchange interaction being insufficient to overcome the large centrifugal barrier, in contrast to the $J^{P}=1/2^{+}$ case, where the interaction is
predominantly repulsive and thus does not support binding.

As shown in Table~\ref{c4}, the predicted binding energies exhibit a strong dependence on the cutoff parameter $\alpha$, particularly for the $I=1/2$, $J^P=7/2^{-}$ and $I=3/2$, $J^P=3/2^{+}$ channels.
In these cases, a very small variation in $\alpha$ leads to a dramatic increase in the binding energy, from only a few MeV to more than one hundred MeV.
\begin{figure}[http]
\begin{center}
\includegraphics[bb=-20 40 750 400, clip, scale=0.42]{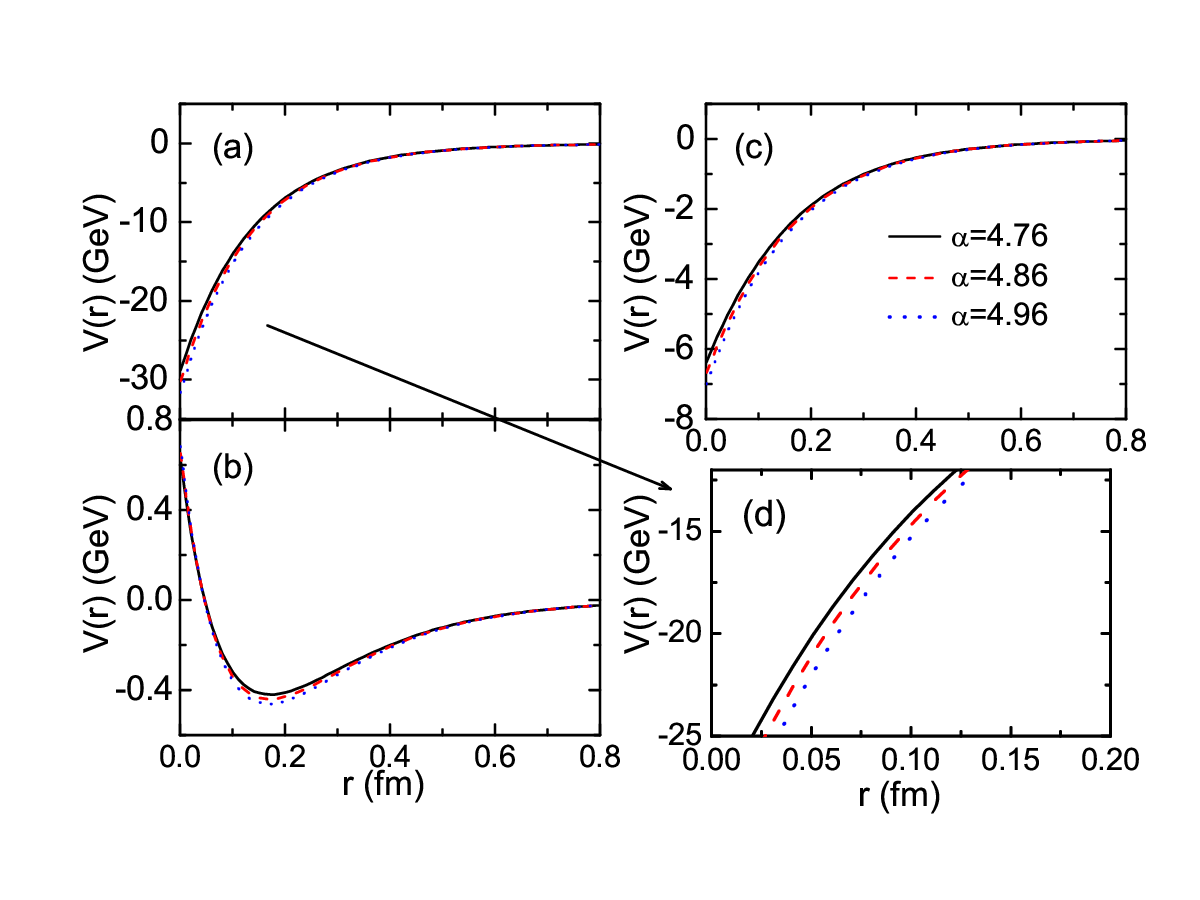}
\caption{Interaction potentials in the $I=1/2$, $J^P=7/2^{-}$ channel. Panels (a), (c), and (b) show the interaction potentials for the $^{4}D_{7/2}$ partial wave, the $^{6}D_{7/2}$
partial wave, and the coupled-channel interaction potential between these two partial waves, respectively. Panel (d) presents an enlarged view of the $^{4}D_{7/2}$ interaction potential
shown in panel (a) in the region $r=0.02$--$0.13~\mathrm{fm}$.}\label{fig:shi-72-minus}
\end{center}
\end{figure}
The main reason is that the interaction potentials in these channels vary dramatically with the cutoff parameter $\alpha$. As an illustrative example, we consider the $I=1/2$, $J^P=7/2^{-}$ channel in detail.
The corresponding interaction potentials for $\alpha$ ranging from 4.76 to 4.96 are presented in Fig.~\ref{fig:shi-72-minus}. As shown in the figure, both the diagonal and off-diagonal interaction potentials are
strongly attractive. Such strong attraction is essential for overcoming the large centrifugal barrier associated with the $D$ wave and thereby generating a bound state. Meanwhile, all of these interaction
potentials exhibit a pronounced dependence on $\alpha$.  Taking the $^{4}D_{7/2}$ partial-wave potential as an example, at $r=0.321~\mathrm{fm}$ the interaction potential increases from $2.9822~\mathrm{GeV}$
for $\alpha=4.76$ to $3.0648~\mathrm{GeV}$ for $\alpha=4.86$, corresponding to a difference of $0.0826~\mathrm{GeV}$. Such a significant variation in the interaction potential can lead to a dramatic change in
the predicted binding energy.

\section*{SUMMARY}
Motivated by recent progress in exotic hadron spectroscopy and the molecular-state interpretation of near-threshold structures, we systematically investigate possible $K^{*}\Sigma^{*}$ molecular states within the one-boson-exchange model by solving the Schr\"odinger equation using the Gaussian expansion method. We focus on states with total spin-parity ranging from $J^{P}=1/2^{\pm}$ to $J^{P}=11/2^{+}$, including higher partial waves and their coupling effects. Both $N$-like ($I=1/2$) and $\Delta$-like ($I=3/2$) isospin configurations are considered.

Our results show that in the $I=3/2$ channel with $J^{P}=7/2^{-}$, a bound state is obtained, which can be associated with the experimentally observed $\Delta(2200)$ resonance as a $D$-wave dominated molecular state. In the $I=1/2$ channel with $J^{P}=9/2^{-}$, a bound state is also found, which may be related to the $N(2250)$ state. In addition, bound-state solutions are obtained in several other $I(J^{P})$ channels.  For $J^{P}=3/2^{\pm},\, 5/2^{\pm},\, 7/2^{+}$, and $9/2^{+}$, molecular states can be formed in both $I=1/2$ and $I=3/2$ channels. For $J^{P}=1/2^{+}$ and $J^{P}=11/2^{+}$, bound states appear only in the $I=3/2$ channel. However, no bound state is found for $J^{P}=1/2^{+}$ in either isospin channel.

It is worth emphasizing that a more precise determination of the spin-parity and decay properties of the $\Delta(2200)$ and $N(2250)$ states in future experiments would be crucial for testing the molecular interpretation proposed in this work. We therefore encourage experimental studies at facilities such as JLab, J-PARC, and PANDA to focus on the relevant decay channels of these resonances, which may provide further insight into their internal structures.

\section*{Acknowledgments}
This work was supported by the National Natural Science Foundation of China under Grant No.12005177.  Yin. Huang also acknowledges the support from the Fundamental Research Funds for the Central Universities under Grant No. 2682026TPY011.

%

\begin{thebibliography}{23}%
\bibitem{ParticleDataGroup:2024cfk}
S.~Navas \textit{et al.} [Particle Data Group],
Phys. Rev. D \textbf{110}, 030001 (2024).




\bibitem{Godfrey:1985xj}
S.~Godfrey and N.~Isgur,
Phys. Rev. D \textbf{32} (1985), 189-231.


\bibitem{Capstick:1986ter}
S.~Capstick and N.~Isgur,
Phys. Rev. D \textbf{34} (1986), 2809-2835.



\bibitem{Guo:2017jvc}
F.~K.~Guo, C.~Hanhart, U.~G.~Mei{\ss}ner, Q.~Wang, Q.~Zhao and B.~S.~Zou,
Rev. Mod. Phys. \textbf{90} (2018), 015004
[erratum: Rev. Mod. Phys. \textbf{94} (2022), 029901].




\bibitem{Belle:2003nnu}
S.~K.~Choi \textit{et al.} [Belle],
Phys. Rev. Lett. \textbf{91}, 262001 (2003).



\bibitem{Brambilla:2019esw}
N.~Brambilla, S.~Eidelman, C.~Hanhart, A.~Nefediev, C.~P.~Shen, C.~E.~Thomas, A.~Vairo and C.~Z.~Yuan,
Phys. Rept. \textbf{873}, 1-154 (2020).


\bibitem{Chen:2022asf}
H.~X.~Chen, W.~Chen, X.~Liu, Y.~R.~Liu and S.~L.~Zhu,
Rept. Prog. Phys. \textbf{86}, 026201 (2023).



\bibitem{Meng:2022ozq}
L.~Meng, B.~Wang, G.~J.~Wang and S.~L.~Zhu,
Phys. Rept. \textbf{1019}, 1-149 (2023).



\bibitem{LHCb:2015yax}
R.~Aaij \textit{et al.} [LHCb],
Phys. Rev. Lett. \textbf{115}, 072001 (2015).


\bibitem{LHCb:2016ztz}
R.~Aaij \textit{et al.} [LHCb],
Phys. Rev. Lett. \textbf{117}, 082002 (2016).


\bibitem{LHCb:2016lve}
R.~Aaij \textit{et al.} [LHCb],
Phys. Rev. Lett. \textbf{117},082003 (2016).


\bibitem{LHCb:2019kea}
R.~Aaij \textit{et al.} [LHCb],
Phys. Rev. Lett. \textbf{122}, 222001 (2019).


\bibitem{LHCb:2020jpq}
R.~Aaij \textit{et al.} [LHCb],
Sci. Bull. \textbf{66}, 1278-1287 (2021).


\bibitem{LHCb:2022ogu}
R.~Aaij \textit{et al.} [LHCb],
Phys. Rev. Lett. \textbf{131}, 031901 (2023).


\bibitem{Chen:2019bip}
H.~X.~Chen, W.~Chen and S.~L.~Zhu,
Phys. Rev. D \textbf{100}, 051501 (2019).

\bibitem{Guo:2019fdo}
F.~K.~Guo, H.~J.~Jing, U.~G.~Mei\ss{}ner and S.~Sakai,
Phys. Rev. D \textbf{99},091501 (2019).


\bibitem{Xiao:2019aya}
C.~W.~Xiao, J.~Nieves and E.~Oset,
Phys. Rev. D \textbf{100},014021 (2019).


\bibitem{He:2019ify}
J.~He,
Eur. Phys. J. C \textbf{79},393 (2019).


\bibitem{Xiao:2019mvs}
C.~J.~Xiao, Y.~Huang, Y.~B.~Dong, L.~S.~Geng and D.~Y.~Chen,
Phys. Rev. D \textbf{100},014022 (2019).


\bibitem{Roca:2015dva}
L.~Roca, J.~Nieves and E.~Oset,
Phys. Rev. D \textbf{92}, 094003 (2015).


\bibitem{Chen:2015moa}
H.~X.~Chen, W.~Chen, X.~Liu, T.~G.~Steele and S.~L.~Zhu,
Phys. Rev. Lett. \textbf{115},172001 (2015).


\bibitem{Chen:2015loa}
R.~Chen, X.~Liu, X.~Q.~Li and S.~L.~Zhu,
Phys. Rev. Lett. \textbf{115},132002 (2015).


\bibitem{Yang:2015bmv}
G.~Yang and J.~Ping,
Phys. Rev. D \textbf{95}, 014010 (2017).


\bibitem{Huang:2015uda}
H.~Huang, C.~Deng, J.~Ping and F.~Wang,
Eur. Phys. J. C \textbf{76}, 624 (2016).


\bibitem{Du:2019pij}
M.~L.~Du, V.~Baru, F.~K.~Guo, C.~Hanhart, U.~G.~Mei\ss{}ner, J.~A.~Oller and Q.~Wang,
Phys. Rev. Lett. \textbf{124} (2020), 072001.



\bibitem{Isgur:1991wq}
N.~Isgur and M.~B.~Wise,
Phys. Rev. Lett. \textbf{66} (1991), 1130-1133.


\bibitem{Huang:2024asn}
Y.~Huang and X.~Chen,
Phys. Lett. B \textbf{868} (2025), 139801.



\bibitem{Oset:1997it}
E.~Oset and A.~Ramos,
Nucl. Phys. A \textbf{635} (1998), 99-120.


\bibitem{Oller:1997ti}
J.~A.~Oller and E.~Oset,
Nucl. Phys. A \textbf{620} (1997), 438-456
[erratum: Nucl. Phys. A \textbf{652} (1999), 407-409].



\bibitem{Wu:2011yx}
J.~J.~Wu, X.~H.~Liu, Q.~Zhao and B.~S.~Zou,
Phys. Rev. Lett. \textbf{108} (2012), 081803.



\bibitem{Roca:2005nm}
L.~Roca, E.~Oset and J.~Singh,
Phys. Rev. D \textbf{72} (2005), 014002.



\bibitem{He:2017aps}
J.~He,
Phys. Rev. D \textbf{95} (2017), 074031.


\bibitem{Sekihara:2015qqa}
T.~Sekihara,
PTEP \textbf{2015} (2015), 091D01.


\bibitem{Khemchandani:2016ftn}
K.~P.~Khemchandani, A.~Mart{\'\i}nez Torres, A.~Hosaka, H.~Nagahiro, F.~S.~Navarra and M.~Nielsen,
Phys. Rev. D \textbf{97} (2018), 034005.


\bibitem{Gamermann:2011mq}
D.~Gamermann, C.~Garcia-Recio, J.~Nieves and L.~L.~Salcedo,
Phys. Rev. D \textbf{84} (2011), 056017.


\bibitem{Miyahara:2016yyh}
K.~Miyahara, T.~Hyodo, M.~Oka, J.~Nieves and E.~Oset,
Phys. Rev. C \textbf{95} (2017), 035212.


\bibitem{Hei:2023eqz}
H.~Hei and Y.~Huang,
Phys. Rev. D \textbf{109} (2024), 016029.



\bibitem{Huang:2020taj}
Y.~Huang and L.~Geng,
Eur. Phys. J. C \textbf{80} (2020), 837.


\bibitem{Huang:2021ahp}
Y.~Huang, F.~Yang and H.~Zhu,
Chin. Phys. C \textbf{45} (2021), 073112.


\bibitem{Yan:2024usf}
Y.~Yan, Q.~Huang, X.~Zhu, H.~Huang and J.~Ping,
Phys. Rev. D \textbf{110} (2024), 014021.



\bibitem{Huang:2018ehi}
H.~Huang, X.~Zhu and J.~Ping,
Phys. Rev. D \textbf{97} (2018), 094019.


\bibitem{Wang:2023eng}
B.~Wang, K.~Chen, L.~Meng and S.~L.~Zhu,
Phys. Rev. D \textbf{109} (2024), 074035.


\bibitem{Yang:2022uot}
P.~Yang and W.~Chen,
Chin. Phys. C \textbf{47} (2023), 013105.


\bibitem{Sarkar:2010saz}
S.~Sarkar, B.~X.~Sun, E.~Oset and M.~J.~Vicente Vacas,
Eur. Phys. J. A \textbf{44} (2010), 431-443.



\bibitem{Hunt:2018wqz}
B.~C.~Hunt and D.~M.~Manley,
Phys. Rev. C \textbf{99} (2019), 055205.



\bibitem{Aachen-Berlin-CERN-London-Vienna:1969bau}
J.~Bartsch \textit{et al.} [Aachen-Berlin-CERN-London-Vienna],
Phys. Lett. B \textbf{28} (1969), 439-442.




\bibitem{Goldwasser:1970fk}
E.~L.~Goldwasser and P.~F.~Schultz,
Phys. Rev. D \textbf{1} (1970), 1960-1966.



\bibitem{Hassall:1981fs}
J.~K.~Hassall, R.~E.~Ansorge, J.~R.~Carter, W.~W.~Neale, J.~G.~Rushbrooke, D.~R.~Ward, B.~Y.~Oh, M.~Pratap, G.~A.~Smith and J.~Whitmore,
Nucl. Phys. B \textbf{189} (1981), 397-420.


\bibitem{Jenkins:1983pm}
C.~M.~Jenkins, J.~R.~Albright, R.~N.~Diamond, H.~C.~Fenker, J.~H.~Goldman, S.~Hagopian, V.~Hagopian, W.~Morris, L.~Kirsch and R.~Poster, \textit{et al.}
Phys. Rev. Lett. \textbf{51} (1983), 951-954.



\bibitem{Biagi:1986zj}
S.~F.~Biagi, M.~Bourquin, R.~M.~Brown, H.~J.~Burckhart, P.~Extermann, M.~Gailloud, C.~N.~P.~Gee, W.~M.~Gibson, P.~Jacot-Guillarmod and J.~Perrier, \textit{et al.}
Z. Phys. C \textbf{34} (1987), 15.


\bibitem{Oset:2002sh}
E.~Oset, J.~R.~Pelaez and L.~Roca,
Phys. Rev. D \textbf{67} (2003), 073013.



\bibitem{Gonzalez:2008pv}
P.~Gonzalez, E.~Oset and J.~Vijande,
Phys. Rev. C \textbf{79} (2009), 025209.



\bibitem{Holmberg:2018dtv}
M.~Holmberg and S.~Leupold,
Eur. Phys. J. A \textbf{54} (2018), 103.


\bibitem{Lutz:2001yb}
M.~F.~M.~Lutz and E.~E.~Kolomeitsev,
Nucl. Phys. A \textbf{700} (2002), 193-308.


\bibitem{Copeland:2020ljp}
P.~M.~Copeland, C.~R.~Ji and W.~Melnitchouk,
Phys. Rev. D \textbf{103} (2021), 094019.


\bibitem{Pascalutsa:2006up}
V.~Pascalutsa, M.~Vanderhaeghen and S.~N.~Yang,
Phys. Rept. \textbf{437} (2007), 125-232.


\bibitem{Ledwig:2011cx}
T.~Ledwig, J.~Martin-Camalich, V.~Pascalutsa and M.~Vanderhaeghen,
Phys. Rev. D \textbf{85} (2012), 034013.


\bibitem{Semke:2005sn}
A.~Semke and M.~F.~M.~Lutz,
Nucl. Phys. A \textbf{778} (2006), 153-180.


\bibitem{Dashen:1993as}
R.~F.~Dashen and A.~V.~Manohar,
Phys. Lett. B \textbf{315} (1993), 425-430.


\bibitem{Matsuyama:2006rp}
A.~Matsuyama, T.~Sato and T.~S.~H.~Lee,
Phys. Rept. \textbf{439} (2007), 193-253.


\bibitem{Hiyama:2003cu}
E.~Hiyama, Y.~Kino and M.~Kamimura,
Prog. Part. Nucl. Phys. \textbf{51} (2003), 223-307.


\bibitem{Dong:2017rmg}
Y.~Dong, A.~Faessler, T.~Gutsche, Q.~L{\"u} and V.~E.~Lyubovitskij,
Phys. Rev. D \textbf{96} (2017), 074027.

\bibitem{Liu:2006df}
X.~Liu, B.~Zhang and S.~L.~Zhu,
Phys. Lett. B \textbf{645} (2007), 185-188.

\bibitem{Xu:2015qqa}
H.~Xu, J.~J.~Xie and X.~Liu,
Eur. Phys. J. C \textbf{76} (2016), 192.

\bibitem{Chen:2012nva}
D.~Y.~Chen, X.~Liu and T.~Matsuki,
Phys. Rev. D \textbf{87} (2013), 054006.


\bibitem{Jian:2022rln}
Z.~Y.~Jian, H.~Q.~Zhu, F.~Yang, Q.~H.~Chen, Y.~Huang and J.~He,
Eur. Phys. J. A \textbf{59} (2023), 262.



\bibitem{Machleidt:1987hj}
R.~Machleidt, K.~Holinde and C.~Elster,
Phys. Rept. \textbf{149} (1987), 1-89.


\bibitem{Weinberg:1990rz}
S.~Weinberg,
Phys. Lett. B \textbf{251} (1990), 288-292.





\end{thebibliography}
\end{document}